\documentclass[usenatbib]{mn2e}

\usepackage{graphics}
\begin{document}

\title[Chemical evolution and galaxy formation]{Chemical signatures of
  formation processes in the stellar populations of simulated
  galaxies}

\author[Tissera et al. ]{Patricia B. Tissera$^{1,2}$,
 Simon D.M. White$^{2}$, Cecilia Scannapieco$^{3}$\\
$^1$ Instituto de Astronom\'{\i}a y F\'{\i}sica del Espacio, Casilla de Correos 67, Suc. 28, 1428, Buenos Aires, Argentina. Conicet - UBA \\
$^2$ Max-Planck Institute for Astrophysics, Karl-Schwarzchild Str. 1, D85748, Garching, Germany\\
$^3$ Leibniz-Institute for Astrophysics Potsdam (AIP), An der Sternwarte 16, D-14482, Postdam, Germany\\
}

\maketitle

\begin{abstract}
We study the chemical properties of the stellar populations in eight
simulations of the formation of Milky-Way mass galaxies in a
$\Lambda$CDM Universe. Our simulations include metal-dependent cooling
and an explicitly multiphase treatment of the effects on the gas of
cooling, enrichment and supernova feedback.  We search for
correlations between formation history and chemical abundance
patterns.  Differing contributions to spheroids and discs from {\it in
  situ} star formation and from accreted populations are reflected in
differing chemical properties.  Discs have younger stellar
populations, with most stars forming {\it in situ} and with low
$\alpha$-enhancement from gas which  never participated in a galactic
outflow.  Up to 15 per cent of disc stars can come from accreted
satellites. These tend to be $\alpha$-enhanced, older and to have
larger velocity dispersions than the {\it in situ} population.  Inner
spheroids have old, metal-rich and $\alpha$-enhanced stars which
formed primarily {\it in situ}, more than $40 $ per cent from material
recycled through earlier galactic winds.  Few accreted stars are found
in the inner spheroid unless a major merger occurred recently. Such
stars are older, more metal-poor and more $\alpha$-enhanced than the
{\it in situ} population. Stellar haloes tend to have low metallicity
and high $\alpha$-enhancement. The outer haloes are made primarily of
accreted stars. Their mean metallicity and $\alpha$-enhancement
reflect the masses of the disrupted satellites where they formed: more
massive satellites typically have higher [Fe/H] and lower
[$\alpha$/Fe].  Surviving satellites have distinctive chemical
patterns which reflect their extended, bursty star formation
histories. These produce lower $\alpha$-enhancement at given
metallicity than in the main galaxy, in agreement with observed trends
in the Milky Way.

\end{abstract}

\section{Introduction}
 
According to the current cosmological paradigm, galaxy formation
occurred through the condensation of gas in the cores of a
hierarchically growing population of dark haloes. This strongly
nonlinear process can be followed in detail only with sophisticated
numerical simulations.  Advances in numerical techniques and in
computational power, together with the accumulation of more precise
and more detailed observations of the galaxy population, have enabled
substantional improvements in our understanding of cosmic structure
formation.  Nevertheless important open questions remain, in
particular on galactic scales \citep{Mo2010}. Many of these issues are
related to the evolution of the baryons. A variety of physical
processes, acting on different scales and with differing efficiency,
are believed to play a role in determining the properties of galaxies.
Direct simulation of galaxy formation in a $\Lambda$CDM context has
evolved greatly in recent years
\citep[e.g.][]{sprin2005,scan06,gove2007,brooks2007,stin2006,sijacki2007, scan09,arge2009,
schaye2010, arge2010,brook2011a} with improved descriptions of cooling, star
formation, black hole formation, and luminous, mechanical and material
feedback from star-forming regions and active galactic
nuclei. Nevertheless, most aspects of this critical baryonic physics
are still described by schematic, phenomenological sub-grid models,
since it is not yet possible to achieve the small-scale resolution
needed to simulate them {\it ab initio}.

Confronting such simulations with observation is a key route to
improving the models and developing a better understanding of galaxy
growth in its cosmological context.  Chemical patterns can be powerful
tools for advancing this programme. The elements are synthesized in
stars and ejected into the interstellar medium (ISM) through stellar
winds and supernovae (SNe). Different SN types release their
nucleosynthesis products on different timescales, imprinting
characteristic chemical patterns on the stellar populations which form
from the gas they enrich \citep[as observed in the Milky Way, see
  e.g][]{freeman,wyse2010}. SN products are mixed with other
interstellar gas by internal mechanisms, or may even be thrown out of
the galaxy as part of a wind. Environmental effects such as
interactions with other galaxies or ram pressure stripping as a result
of motion through the intergalactic medium, also play a role by mixing
new heavy elements into the intergalactic medium. Galaxy mergers
induce major starbursts which substantially modify the abundance
patterns in the ISM \citep[e.g.][Perez et al. submitted]{perez2006,dimat2009,rupke2010}.  Even
minor mergers with infalling satellites modify metal abundance
patterns by contributing stars and gas with distinctive abundances and
by triggering new star formation activity
\citep[e.g.][]{bh96,mh96,tissera2000}.

The possible contribution of disrupted satellites to the stellar
populations of thin discs \citep{abadi2003}, thick discs
\citep{huang97} and bulges \citep{rahimi2010} has been much debated.  Some simulations
have suggested that stellar haloes may have formed primarily through
satellite disruption \citep[e.g.][]{john2008,cooper2010}, but there
have also been claims for a dual formation route, depending on the
early merging history of each individual galaxy
\citep[e.g.][]{zolo2009, zolo2010}; this may be supported by abundance
patterns in the stellar halo and satellites of the Milky Way
\citep{venn2004}.  Different evolutionary paths are expected to leave
their fingerprints on the chemical and dynamical properties of the
stellar populations, and analysing such fingerprints should help us to
learn more about how galaxies formed.

Much work has been devoted to understanding if chemical patterns can
indeed be related to specific events during galaxy formation
\citep[e.g.][]{molla1995,chiap1997}. Within the current hierarchical
paradigm, the nonlinearity and complexity of structure growth require
chemical and structural evolution to be treated simultaneously and
consistently if reliable conclusions are to be drawn.  In the last
decade a variety of treatments of chemical evolution have been
implemented in the cosmological hydrodynamics codes used to simulate
galaxy formation \citep[e.g.][]{rait1996, mosc2001,lia2002,
  kg2003,koba2007,wier2009,mart2008}. These typically adopt the
standard metal production assumptions from analytic models of chemical
evolution \citep[e.g.][]{tins1979, matt1986} but follow the formation
and transport of elements among the various components of the system
in a dynamically consistent way. Such codes have already given a
number of encouraging results, reproducing general trends observed in
the galaxy population such as the mass-metallicity relation, for
example \citep{gove2007,koba2007}.

The simulations analysed in the current paper follow the formation of
galaxies within a set of eight Milky-Way-mass haloes from the Aquarius
Project \citep{sprin2008}.  The resulting central galaxies have been
extensively studied in papers by \citet{scan09}, \citet{tissera2010},
\cite{scan10} and \cite{scan11}.  These studies highlighted the
difficulty in forming spirals similar to the Milky Way within the
$\Lambda$CDM cosmology  (but see \citet{guedes2011} who were able to produce an $L^*$ disk galaxy with a low $i$-band bulge-to-disk ratio in a single
high-resolution simulation which assumed a higher surface density 
threshold for star formation than earlier simulations like our own.).  
\citet{scan09} found that although most of
the simulated galaxies had a substantial disc at some stage during
their evolution, these discs are often disrupted by interactions with
satellites or destabilized by misaligned inflowing cold gas.  These
processes feed the growth of stellar spheroids, which end up as the
dominant components by mass in all the simulations.  Thus, at $z=0$
none of our simulated haloes hosts a spiral with a disc-to-total mass ratio similar to that of 
the Milky Way. In spite of this, the final discs and spheroids
show general dynamical properties which are quite consistent with
those of observed spiral galaxies.  While the spheroids are dominated
by old, dynamically hot stars  (and tend to rotate more slowly than observed bulges), the discs have the youngest stars, they
formed typically from the inside out, and they are dynamically cold.
 Residual cold gas forms a well-behaved disc in equilibrium within its potential
well \citep{scan11}.  The fact that these
two components show distinct star formation histories and distinct
dynamical properties prompted us to explore their chemical properties,
seeking chemical signatures which can be related to specific details
of their formation histories.

The simulations studied here use a specially enhanced version of
{\small GADGET-3} which is itself an optimisation of {\small GADGET-2}
\citep{sprin2005}. Our version includes a chemical and energy feedback
model which is consistently coupled to a unique multiphase
representation of the interstellar medium (ISM) which
 allows collocation, material exchange, and relative
motion between the different phases. This scheme allows us to follow
the dynamical and chemical properties of baryons throughout the galaxy
formation process. Dense and cold gas is transformed into stars which
can explode as SN. These inject entropy into some of the remaining
cold gas, converting it to the hot diffuse phase and driving powerful,
mass-loaded winds. These thermodynamic changes are generated on a
particle-by-particle basis in a way which is consistent with local
astrophysical conditions. As a result star-forming regions drive winds
without the need to temporarily suspend cooling, to impart large kicks
to particles, or to suppress hydrodynamical interactions with
surrounding particles.
 This naturally results in self-regulated winds which are
adapted to the depth of the potential well in which star formation is
occurring \citep{scan06,scan09,sawala2010,sawala2011,dero10}.  In this
framework, gas is progressively enriched as subsequent stellar
generations release new elements into the ISM, so that the chemical
structure of each galaxy is intimately related to the formation paths
of its various components.  In this paper, we focus on this relation
between chemical properties and assembly history. In a subsequent
paper, we will analyse how gas flows lead to abundance gradients
(Tissera et al. 2011 in preparation).

In Section 2 we describe our simulations and the main aspects of our
chemical model. We also include a subsection explaining the criteria
we adopt to separate our galaxies into their dynamical components.  In
Section 3 we study the chemical properties of these components,
together with their formation histories. Section 4 studies the
chemical properties of stars formed {\it in situ} and contrasts them
with those of stars accreted from tidally disrupted satellites.
Section 5 compares the mean chemical abundances of stars from
disrupted satellites with those of stars in surviving satellites.  We
summarize our findings in the concluding section.

\section{The simulations}

We have analysed galaxy formation within a set of eight haloes
selected and simulated at high resolution as part of the Aquarius
Project \citep{sprin2008}.  These objects were identified in a
lower-resolution version of the Millennium-II Simulation
\citep{boylan2009}, a cosmological simulation of a cubic region $100
\ { \rm Mpc \ h^{-1}}$ on a side, imposing a mild isolation criterion
and requiring a mass similar to that of the Milky Way's halo at
$z=0$. In the original project these haloes were simulated at a
variety of resolutions following the dark matter only.  For the
simulations discussed here, gas particles were added to the initial
conditions on a cubic mesh displaced by half the interparticle
separation from the dark matter particle mesh. Details can be found in
\citet{scan09}.
 
All these simulations assumed a $\Lambda$CDM cosmogony with
$\Omega_{\rm m}=0.25,\Omega_{\Lambda}=0.75, \Omega_{\rm b}=0.04,
\sigma_{8}=0.9, n_{s}=1$ and $H_0 = 100 \ h \ { \rm km s^{-1}
  Mpc^{-1}}$ with $h =0.73$. Our galaxy formation simulations were
carried out with maximum gravitational softenings in the range
$\epsilon_{G} =0.5 - 1$ kpc $h^{-1}$, dark matter particle masses of
order $10^{6}{\rm M_\odot }h^{-1}$ and initial gas particle masses of
order $2 \times 10^{5}{\rm M_\odot} h^{-1}$. Since their virial masses
are in the range $5$ to $11 \times 10^{11} {\rm M_\odot} h^{-1}$, the
final systems all have about a million particles in total within the
virial radius\footnote{We define the virial radius ($r_{200}$) as the
  radius within which the mean mass density of all components is $200$
  times the critical value. The virial mass is then the total mass
  within this radius.}.  In Table ~\ref{tab1} we summarize their
principal characteristics. Our naming convention (Aq-A-5, etc.)
follows that of earlier papers. The capital letter (A to H) designates
the particular halo considered, whereas the number (1 to 5) indicates
the resolution level of the simulation according to the original
Aquarius convention. All simulations studied in this paper are at
level 5 the lowest of the original Aquarius resolutions. Further
details can be found in \citet{scan09}.

\citet{scan09} and \citet{scan11} studied the formation of the disc
and the spheroidal components of these systems and their dynamical and
photometric properties, while \citet{tissera2010} analysed the reponse
of their dark matter haloes to the galaxy formation process. The main
galaxies have a variety of morphologies with varying bulge-to-disc
ratios and strong bars in several cases. Nevertheless, all of them are
bulge dominated, and one (Aq-F-5) has no detectable disc at all at
$z=0$.  In this paper, we focus on the chemical properties and the
assembly histories of the various stellar components of the galaxies.

\subsection{The galaxy formation simulation code}

The version of {\small GADGET-3} used to carry out these simulations
has been extended to include an explicit multiphase model for the gas
component which allows a cold, dense, possibly star-forming phase and
a hot, diffuse phase to interpenetrate, to move relative to each
other, and to exchange material. This implementation also includes
metal-dependent cooling, star formation, and chemical and thermal
feedback into both gas phases. as described in \citet{scan05} and
\cite{scan06}.  This scheme has been successful in reproducing the
star formation activity of galaxies during both quiescent and
starburst phases, and can drive substantial mass-loaded galactic winds
from star-forming galaxies with a wide range of masses at speeds that
consistently reflect the depth of the individual potential wells
\citep{scan06, scan08, sawala2010, sawala2011, dero10}.  The scheme does not
require any  {\it ad hoc} ``fixes''  (for example, temporary
suppression of radiative cooling and/or of hydrodynamic coupling for
wind particles, or the imposition of discontinuous momentum changes to
``kick-start'' a wind) 
to ensure efficient wind
generation in SPH simulations, and there are no scale-dependent
parameters.
 As a result, it is particularly well suited for the study of galaxy formation in a
cosmological context where galaxies with a wide range of properties
form simultaneously.  The version used here includes chemical
enrichment by Type II and Type Ia supernovae (SNII and SNIa,
respectively) as we describe next.

\subsubsection{The Chemical Model}

The chemical evolution model adopted in this work is that developed by
\citet{mosc2001} and adapted by \citet{scan09} to {\small GADGET-3}.
We consider chemical elements which are synthesised in stellar
interiors and then ejected into the ISM by SNII and SNIa events.  A
Salpeter Initial Mass Function is assumed with lower and upper mass
cut-offs at 0.1 ${\rm M_{\odot}}$ and 40 ${\rm M_{\odot}}$,
respectively.  We follow 12 different chemical isotopes: H, $^4$He,
$^{12}$C, $^{16}$O, $^{24}$Mg, $^{28}$Si, $^{56}$Fe, $^{14}$N,
$^{20}$Ne, $^{32}$S, $^{40}$Ca and $^{62}$Zn.  Initially, all baryons
are in the form of gas with primordial abundances, i.e. 76\% H and
24\% $^4$He by mass.

SNII are considered to originate from stars more massive than 8 ${\rm
  M_{\odot}}$, adopting the metal-dependent yields of \citet{WW95}.
We use the metal-and-mass dependent lifetimes given by the fitting formulae of
\citet{rait1996}.
For SNIa, we adopt the W7 model of \citet{thie1993} which assumes that
SNIa events originate from CO white dwarf systems in which mass is
transferred from the secondary to the primary star until the
Chandrasekhar mass is exceeded and an explosion is triggered.  For
simplicity, we assume that the lifetime of the progenitor systems is
distributed uniformly and at random over the range $[0.7, 2]$
Gyr. To calculate the number of SNIa, we adopt an observationally
motivated rate of 0.3 relative to SNII 
 \citep{mosc2001}. 
  
The ejection of chemical elements is grafted onto the SN feedback
model so that chemical elements are distributed to the cold and
the hot gas phases surrounding a given star particle. The fraction of
elements going to each phase is regulated by a free parameter
$\epsilon_{\rm c}$ which also determines the amount of SN energy
received by each phase. While the injection of energy follows two
different paths depending on the thermodynamical properties of the
gas, chemical elements are added to both phases at the time each
SN occurs.  All the simulations of this paper have been run with
$\epsilon_{\rm c}=0.5$. Some tests of the effect of varying this
parameter can be found in \citet{scan06}.

\begin{table*}
  \begin{center}
  \caption{General characteristics of our main galaxies and their
    haloes at $z=0$. The left column shows the halo name.  $r_{\rm 200}$
    and $M_{\rm 200}$ are the virial radius and mass, respectively.
    $N_{\rm dm}$, $N_{\rm gas}$, $N_{\rm hot}$ and $N_*$ are the total
    numbers of dark matter, gas, hot gas and star particles within
    $r_{\rm 200}$, respectively. $r_{\rm opt}$ is the optical radius of the galaxy
    and $M_{\rm *}^{\rm opt}$ and $N_*^{\rm opt}$ are the stellar mass and the
    number of star particles estimated within twice $r_{\rm opt}$.
  }
\label{tab1}
\begin{tabular}{|l|c|c|c|c|c|c|c|c|c|c|c}\hline
 {Halo}   & $r_{\rm 200}$& $M_{\rm 200}$& $M_{\rm hot}$ & $M_{\rm cold}$&$N_{\rm dm}$ & $N_{\rm gas}$ &$N_{\rm hot}$ &$N_*$ 
& $r_{\rm opt}$ &$M_*^{\rm opt}$ &$N_*^{\rm opt}$\\ 
 & $h^{-1}$kpc &$10^{12}{h^{-1}\rm M_{\odot}} $&$10^{10}{h^{-1}\rm M_{\odot}} $ &$10^{10}{h^{-1}\rm M_{\odot}} $&&&&& h$^{-1}$kpc &$10^{10}h^{-1}{\rm M_{\odot}}$ \\\hline
 Aq-A-5  & 169.42 &  1.10 & 4.1&1.1& 529110& 92159&72157&333409  &13.1 & 5.92&277382\\
Aq-B-5  & 132.10 &  0.52  & 1.2&0.6& 435330& 63580&41312&291313  &13.0 & 2.53&244511\\
Aq-C-5  & 173.19 &  1.18  & 3.4&0.6& 681143& 97796&83845&548964  &11.7 & 5.93&432232\\
Aq-D-5  & 170.63 &  1.09  & 1.7&0.2& 599438& 82736&75741&377866  &10.8 & 4.41&287734\\
Aq-E-5  & 149.93 &  0.79  & 2.1&0.6& 554245& 80524&63458&512809  &10.1 & 4.97&406399 \\
Aq-F-5  & 142.74 &  0.67  & 1.6&0.2& 680129& 79345&71124&699052  &10.3 & 5.39&525101 \\
Aq-G-5  & 142.64 &  0.68  & 1.2&0.5& 679177& 74675&52828&387751  &10.3 & 5.63&334108\\
Aq-H-5  & 132.94 &  0.53  & 0.5&0.1& 515392& 23133&20671&546482  &7.60 & 4.73&425680\\
\end{tabular} 
 \end{center}
\vspace{1mm}
\end{table*}

\begin{table*}
  \begin{center}
  \caption{ Properties of the stellar components of the main galaxies at $z=0$.
$M_{\rm disc}$, $M_{\rm cen}$, $M_{\rm Inner}$ and $M_{\rm Outer}$ are the stellar masses of disc, central spheroid, inner halo and outer halo of the main Aquarius 
galaxies  as obtained from the kinematic decomposition. All masses are
given in units of $10^{9}{h^{-1}\rm M_{\odot}} $. 
 $D/T^{1}$ and $D/T^{2}$ are the disc-to-total stellar mass
    considering as total mass only the stellar discs and the central
    spheroids ($T^{1}$) and also the stellar inner haloes $(T^{2})$.
 The last column show  ${\rm D/T^{photo}}$ photometric ratios calculated by \citet{scan10}
using an observationally based approach  (for Aq-F-5 amd Aq-H-5, they could
not  obtain a reliable photometric decomposition). }
\label{tab1}
\begin{tabular}{|l|c|c|c|c|c|c|c}\hline
 {Halo}   & $M_{\rm disc}$& $M_{\rm cen}$& $M_{\rm Inner}$ & $M_{\rm Outer}$ &${\rm D/T^{1}}$ & ${\rm D/T^{2}}$& ${\rm D/T^{photo}}$\\\hline
Aq-A-5  &3.2 &  3.4   & 14.6 & 2.4&0.08&0.06&0.32\\
Aq-B-5  &2.2 &  16.9  & 6.4  & 1.4&0.11&0.09&0.42\\
Aq-C-5  &1.6 &  37.8  & 14.2 & 3.6&0.29&0.23&0.49\\
Aq-D-5  &8.9 &  24.0  & 13.6 & 4.0&0.27&0.19&0.68\\
Aq-E-5  &1.1 &  25.2  & 15.3 & 2.1&0.30&0.21&0.40\\
Aq-F-5  & - &   39.0  & 9.5 & 0.5&---&---& ---\\
Aq-G-5  &6.0 &  16.0  & 7.3 & 0.7&0.27&0.20&0.60\\
Aq-H-5  &1.3 &  15.6  & 22.3 & 2.5&0.07&0.03& ---\\
\end{tabular} 
 \end{center}
\vspace{1mm}
\end{table*}

\begin{table*}
  \begin{center}
  \caption{Chemical properties of the stellar discs. The first column
    gives the encoding name of the halo. $F_{\rm prog}$ and $F_{\rm
      prom}$ are the fractions of stars formed in the progenitor
    systems and of stars which have been part of a galactic wind.
    $\sigma/V$ is the ratio of velocity dispersion to mean circular
    velocity. Age is the mean stellar age in Gyr. [Fe/H] and [O/Fe]
    are median logarithmic abundances relative to solar. Standard
    deviations are included within parentheses. The superscripts
    ``prog'' and ``sat'' denote values corresponding to stars formed
    {\it in situ} and to stars acquired from disrupted satellites,
    respectively.  }
\label{tabdisc}
\begin{tabular}{|l|c|c|c|c|c|c|c|c|c|c}\hline
 {Halo}  & $F_{\rm prog}$& $F_{\rm prom}$&$\sigma/V^{\rm prog}$& $(\sigma/V)^{\rm sat}$& [Fe/H]$^{\rm prog}$&[Fe/H]$^{\rm sat}$&[O/Fe]$^{\rm prog}$ & [O/Fe]$^{\rm sat}$ & Age$^{\rm prog}$ & Age$^{\rm sat}$  \\\hline
Aq-A-5 &0.97&0.08 &0.86&1.13& -1.10(0.43)  &-1.75(1.35)& -0.163(0.133)& 0.343(0.052)& 5.89(7.53) &12.66(0.13)\\
Aq-B-5 &0.91&0.09 &0.67&0.89& -1.25  (0.62)&-1.55(1.03)&-0.174(0.154) & 0.094(0.288)& 4.08(4.19) &10.63(4.92)\\
Aq-C-5 &0.99&0.09 &0.59&0.76& -0.86 (0.33) &-1.37(0.81)&-0.131(0.121) & 0.300(0.113)& 8.69(5.07) &12.39(3.07)\\
Aq-D-5 &0.89&0.15 &0.68&0.77& -0.85 (0.34) &-0.97(0.73)& -0.096(0.114)& 0.192(0.142)& 7.99(6.47) &10.81(1.26)\\
Aq-E-5 &0.62&0.30 &0.54&0.60& -0.95 (0.36) &-0.83(0.58)& -0.049(0.123)& 0.031(0.154)& 8.53(2.98) &11.09(0.79)\\
Aq-F-5 &--- &---  & ---&--- &---           &       --- &            ---&---   & --- & ---\\
Aq-G-5 &0.97&0.12 &0.62&0.73& -0.94(0.38)  &-1.75(1.21)& -0.092(0.109)& 0.076(0.259)& 8.06(7.07) &12.22(2.85)\\
Aq-H-5 &0.84&0.21 &1.08&1.02& -0.90(0.39)  &-1.25(1.08)& -0.137(0.131)& 0.241(0.151)& 7.72(7.03) &12.36(0.59)\\
\end{tabular} 
 \end{center}
\vspace{1mm}
\end{table*}

\begin{table*}
  \begin{center}
  \caption{Chemical properties of the central spheroids. The first
    column gives the encoding name of the halo.  $F_{\rm prog}$ and
    $F_{\rm prom}$ are the fractions of stars formed in the progenitor
    systems and of stars which have been part of a galactic wind.  $\sigma$
    is the mean velocity dispersion (in km s$^{-1}$), while Age is the
    mean stellar age (in Gyr). [Fe/H] and [O/Fe] are median
    logarithmic abundances relative to solar. Standard deviations are
    included within parentheses. The superscripts ``prog'' and ``sat''
    label values corresponding to stars formed {\it in situ} and to
    stars acquired from disrupted satellites, respectively. }
\label{tabbulge}
\begin{tabular}{|l|c|c|c|c|c|c|c|c}\hline
 {Halo} &  $F_{\rm prog}$&$F_{\rm prom}$&[Fe/H]$^{\rm prog}$& [Fe/H]$^{\rm sat}$&[O/Fe]& [O/Fe]$^{\rm sat}$& Age$^{\rm prog}$ & Age$^{\rm sat}$  \\\hline
 Aq-A-5& 0.96& 0.43 &-0.61(0.25)& -1.20(0.83) & 0.023(0.126)& 0.358(0.016)& 11.46(1.67)&12.63(0.09)\\
 Aq-B-5& 0.79& 0.64&-0.66(0.36) &-0.92(0.55) &-0.025(0.111)& 0.159(0.092)& 9.43(1.52) &10.93(1.56)\\
 Aq-C-5& 0.95& 0.44&-0.61(0.34) &-1.12(0.56) & 0.038(0.122)& 0.353(0.033)& 11.82(1.18)&12.52(0.04)\\
 Aq-D-5& 0.90& 0.60&-0.62(0.29) &-0.99(0.63) &-0.003(0.110)& 0.214(0.084)& 10.75(1.39)&11.95(0.20)\\
 Aq-E-5& 0.71& 0.75&-0.40(0.31) &-0.57(0.41) &-0.031(0.108)&0.014(0.109)& 10.56(3.95) &11.53(0.65)\\
 Aq-F-5& 0.55& 0.53&-0.52(0.32) &-0.65(0.50) &-0.061(0.103)& 0.037(0.118)& 10.58(4.62)&10.93(3.02)\\
 Aq-G-5& 0.74& 0.73&-0.58(0.35) &-0.70(0.50) & 0.048(0.112)& 0.122(0.085)& 11.37(2.00)&11.08(1.53)\\
 Aq-H-5& 0.83& 0.78&-0.40(0.33) &-0.90(0.46) &-0.026(0.122)& 0.358(0.028)& 11.62(1.51)&12.45(0.09)\\
\end{tabular} 
 \end{center}
\vspace{1mm}
\end{table*}

    \begin{table*}
  \begin{center}
  \caption{Chemical properties of the inner diffuse stellar  haloes. The first
    column shows the encoding name of the simulations. $F_{\rm prog}$
    and $F_{\rm prom}$ are the fractions of stars formed in the
    progenitor systems and of stars which have been part of a galactic
    wind.  $\sigma$ and Age are the mean velocity dispersion (in km
    s$^{-1}$) and mean age (in Gyr), respectively. [Fe/H] and [O/Fe]
    are median abundances. Standard deviations are included within
    parentheses. The labels ``prog'' and ``sat'' denote the values
    corresponding to stars formed {\it in situ} and to stars acquired
    from satellites. }
\label{tabhalo}
\begin{tabular}{|l|c|c|c|c|c|c|c|c}\hline
 {Halo}  &  $F_{\rm prog}$& $F_{\rm prom}$&  [Fe/H]$^{\rm prog}$& [Fe/H]$^{\rm sat}$&[O/Fe]$^{\rm prog}$& [O/Fe]$^{\rm sat}$& Age$^{\rm prog}$ & Age$^{\rm sat}$   \\\hline
Aq-A-5& 0.75& 0.15& -1.17(0.52)&-1.45(1.07) &0.060(0.145) &0.351(0.054) & 10.40(4-05) &12.60(0.15)\\
Aq-B-5& 0.32& 0.27& -1.37(0.85)&-1.24(0.79) &0.031(0.150) &0.113(0.223) & 10.45(1.94) &10.12(2.33)\\
Aq-C-5& 0.56& 0.26& -1.05(0.81)&-1.20(0.8)  &0.002(0.141) &0.311(0.008) & 11.50(0.92) &12.37(0.40)\\
Aq-D-5& 0.57& 0.27& -1.09(0.54)&-1.03(0.75) &-0.079(0.135)&0.152(0.146) & 9.96(2.67)  &11.47(0.53)\\
Aq-E-5&	0.47& 0.43& -0.84(0.51)&-0.86(0.63) &-0.049(0.139)&0.028(0.154) & 10.29(4.12)  &11.16(0.76)\\
Aq-F-5& 0.25& 0.39& -1.05(0.59)&-0.87(0.68) &-0.002(0.131)&-0.024(0.184)& 9.44(5.59)  &9.76(3.89)\\
Aq-G-5& 0.59& 0.35& -1.03(0.58)&-1.02(0.93) &0.019(0.126) &0.043(0.177) & 11.05(1.89) &10.47(2.13)\\
Aq-H-5& 0.54& 0.46& -0.79(0.49)&-0.87(0.67) &-0.056(0.175)&0.148(0.420) & 10.88(2.10) &11.82(0.88)\\
\end{tabular} 
 \end{center}
\vspace{1mm}
\end{table*}

    \begin{table*}
  \begin{center}
  \caption{Chemical properties of the outer diffuse stellar  haloes. The first
    column shows the encoding name of the simulations.  $F_{\rm prog}$
    and $F_{\rm prom}$ are the fractions of stars formed in the
    progenitor systems and of stars which have been part of a galactic
    wind.  $\sigma$ and Age are the mean velocity dispersion (in km
    s$^{-1}$) and age (Gyr)of the stars. [Fe/H] and [O/Fe] are median
    abundances. Standard deviations are included within parentheses.
    The labels ``prog'' and ``sat'' denote the values corresponding to
    stars formed {\it in situ} and to stars acquired from
    satellites. }
\label{tabhaloex}
\begin{tabular}{|l|c|c|c|c|c|c|c|c}\hline
 {Halo} & $F_{\rm prog}$&$F_{\rm prom}$& [Fe/H]$^{\rm prog}$& [Fe/H]$^{\rm sat}$&[O/Fe]$^{\rm prog}$& [O/Fe]$^{\rm sat}$&  Age$^{\rm prog}$  & Age$^{\rm sat}$   \\\hline
Aq-A-5 &0.35 &0.20&-2.04(1.25) &-1.44(1.31) &0.018(0.168)& 0.145(0.203) & 10.79(3.93) & 11.08(2.14)\\
Aq-B-5 &0.13 &0.12&-1.59(1.39) &-1.70(1.12) &0.019(0.154)& 0.119(0.307) & 10.14(2.43) & 9.28(2.58)\\
Aq-C-5 &0.19 &0.21&-2.04(1.46) &-1.31(1.14) &0.054(0.149)& 0.089(0.205) & 11.54(0.92) & 11.29(1.11)\\
Aq-D-5 &0.19 &0.24& -1.76(1.24)&-1.36(1.07) &0.018(0.160)& 0.100(0.249) & 10.59(2.47) & 11.08(1.22)\\
Aq-E-5 &0.25 &0.20&-1.66(1.17) & 1.39(1.16) &0.019(0.155)& 0.079(0.242) & 10.82(1.52) & 11.32(1.67)\\
Aq-F-5 &0.03 &0.23&-2.11(1.65) &-1.32(1.09) &0.019(0.160)& 0.019 (0.302)& 11.14(6.99) & 9.64(3.41)\\
Aq-G-5 &0.25 &0.05&-1.78(1.44) &-1.85(1.22) &0.018(0.161)& 0.089(0.320) & 10.55(6.45) & 10.54(4.07)\\
Aq-H-5 &0.15 &0.29&-1.73(1.64) &-1.21(1.02) &0.022(0.136)& 0.075(0.222) & 11.80(2.36) & 11.21(1.21)\\
\end{tabular} 
 \end{center}
\vspace{1mm}
\end{table*}


\subsection{The  stellar components of the simulated galaxies}

As shown by \citet{scan09}, the simulated galaxies have a variety of 
morphologies and contain both  spheroidal and  disc components.
Our main contribution to understanding how these galaxies formed 
comes from  linking their dynamical evolution with specific chemical 
patterns in their stellar populations.

\begin{figure*}
\hspace*{-0.2cm}\resizebox{6cm}{!}{\includegraphics{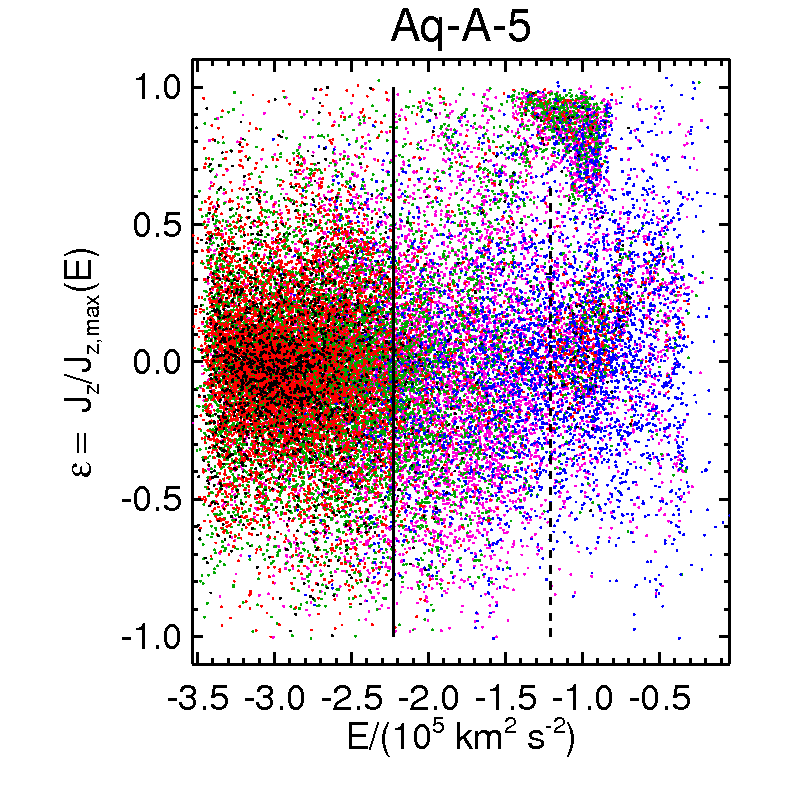}}
\hspace*{-0.5cm}\resizebox{6cm}{!}{\includegraphics{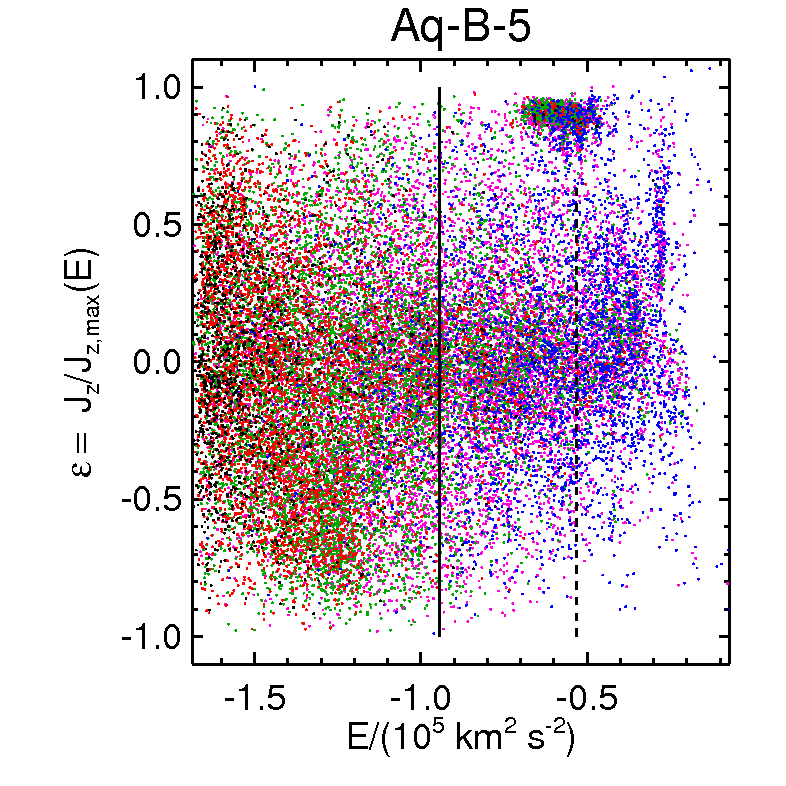}}
\hspace*{-0.5cm}\resizebox{6cm}{!}{\includegraphics{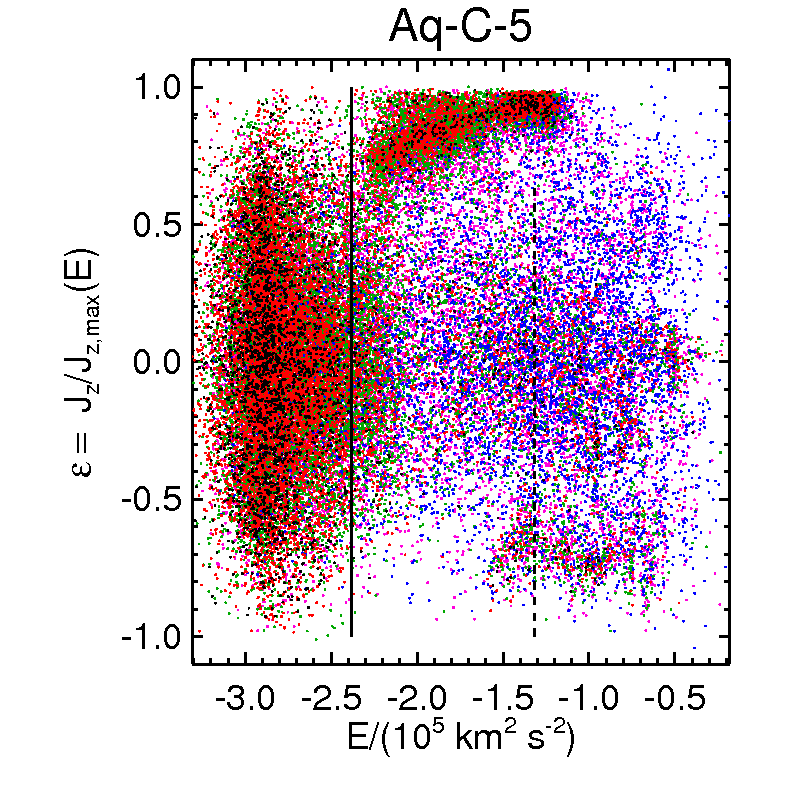}}\\
\hspace*{-0.2cm}\resizebox{6cm}{!}{\includegraphics{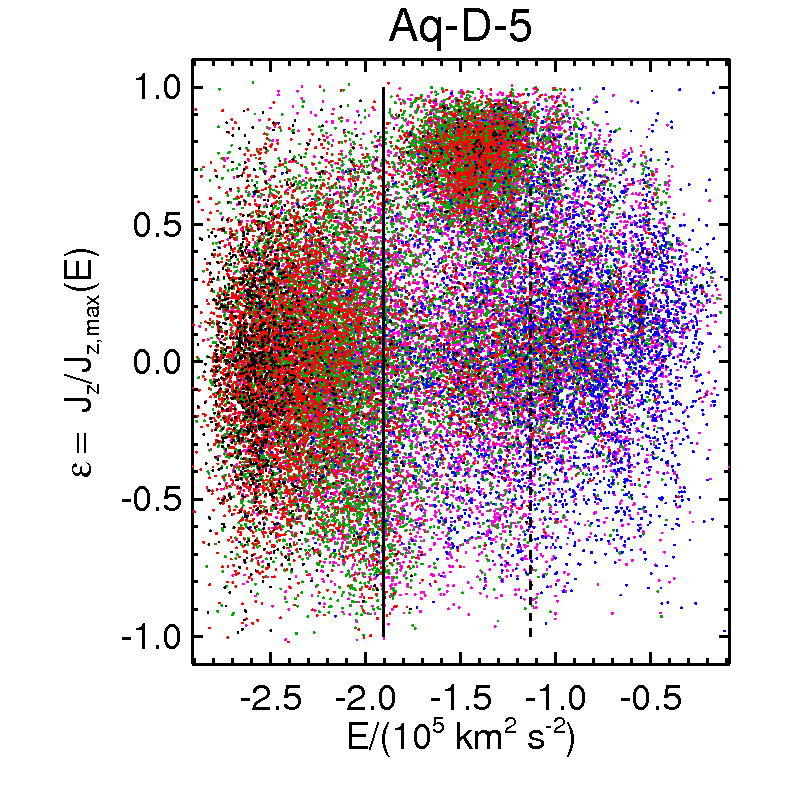}}
\hspace*{-0.5cm}\resizebox{6cm}{!}{\includegraphics{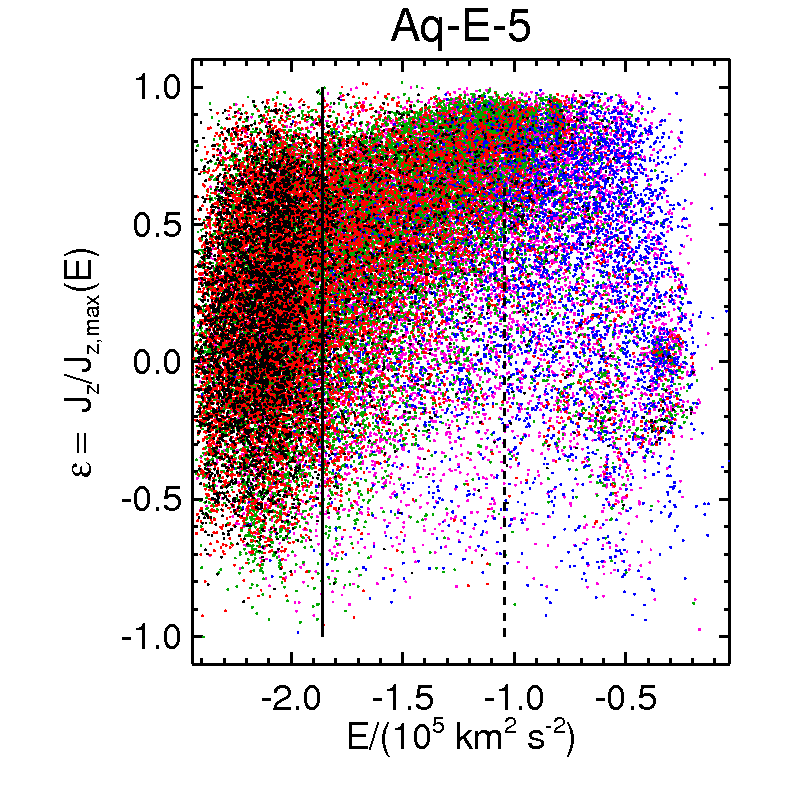}}
\hspace*{-0.5cm}\resizebox{6cm}{!}{\includegraphics{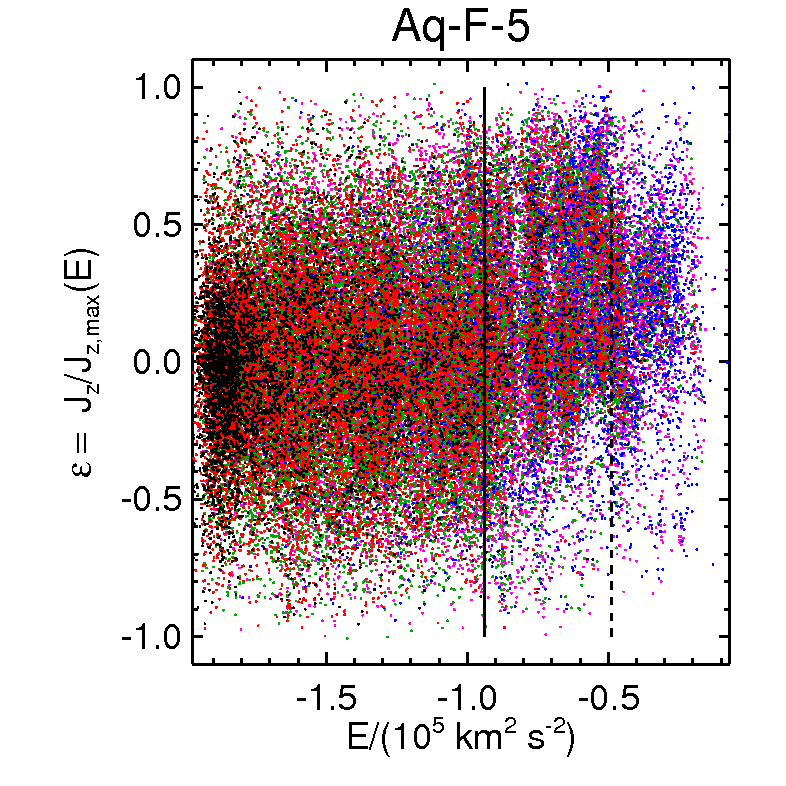}}\\
\hspace*{-0.2cm}\resizebox{6cm}{!}{\includegraphics{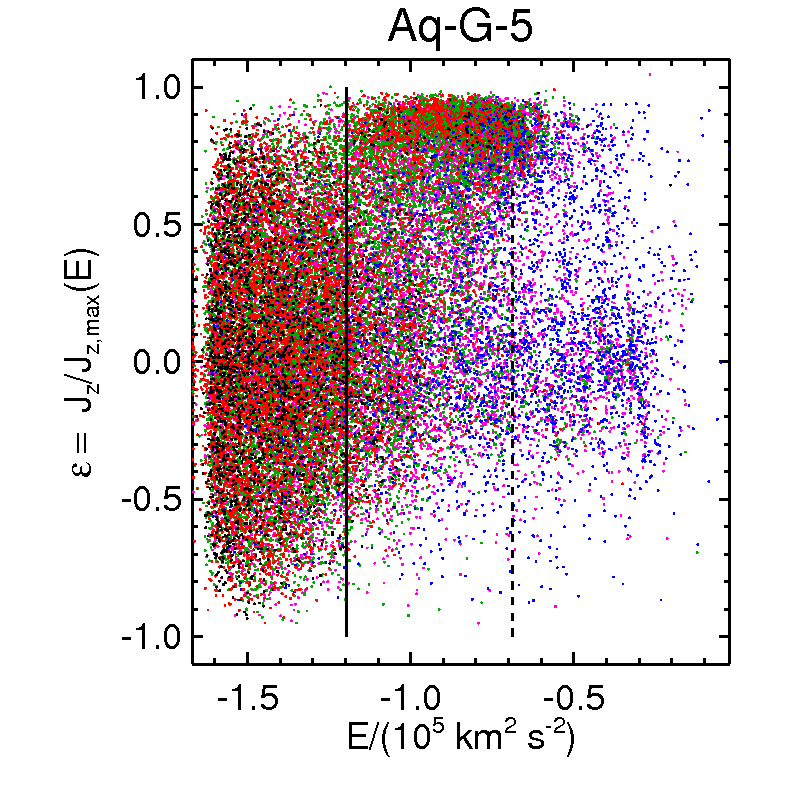}}
\hspace*{-0.5cm}\resizebox{6cm}{!}{\includegraphics{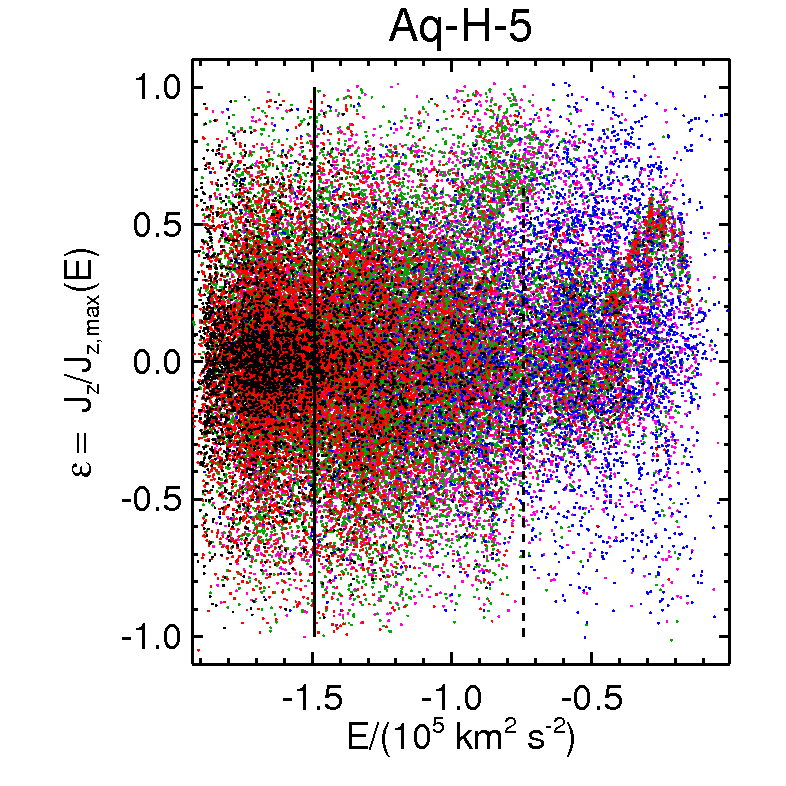}}
\hspace*{-0.5cm}\resizebox{6cm}{!}{\includegraphics{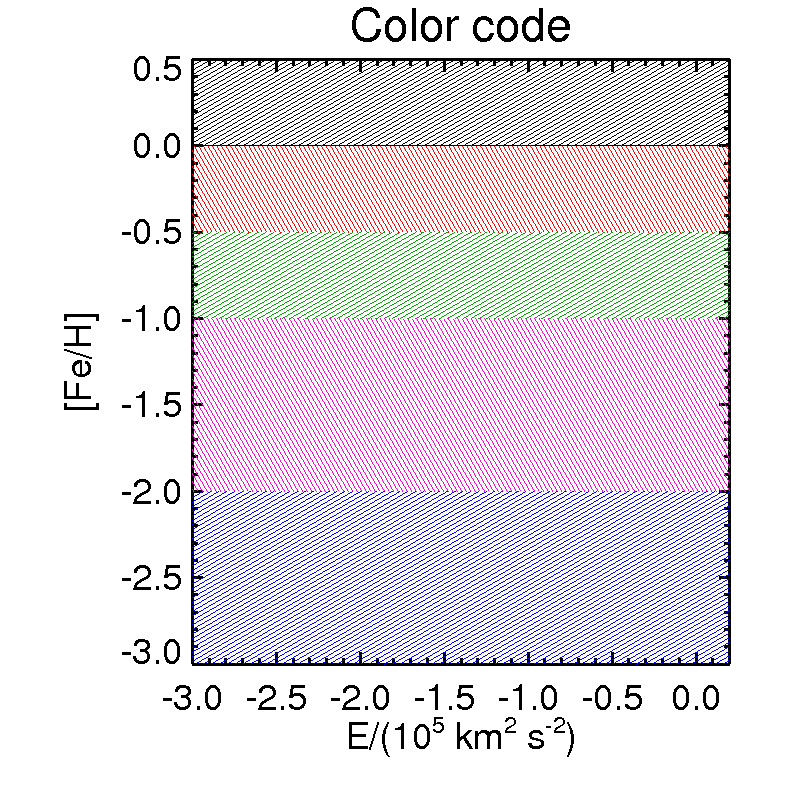}}
\caption{Distribution of $\epsilon=J_z/J_{z,\rm max}(E)$ as a function
  of energy $E$ for our eight Aquarius haloes. The last panel shows
  the color code used to represent the [Fe/H] abundances of star
  particles. Energies have been zero-pointed at the value for a
  circular orbit at the virial radius. The energies which
  separate our three spheroidal components are indicated for galaxy
  system in the relevant panel.  For the sake of clarity, only $10\%$ of the  particles are displayed.}
\label{scatter}
\end{figure*}

Several methods for separating spheroids and discs have been proposed
\citep[e.g.][]{abadi2003,gove2009, scan09}.  In this work we use a
procedure similar to that of \citet{abadi2003}.  We measure
$\epsilon=J_z/J_{z,\rm max}(E)$ for each star, where $J_z$ is the
angular momentum component perpendicular to the disc plane and
$J_{z,\rm max}(E)$ is the maximum $J_z$ over all particles of given
total energy, $E$. A star on a prograde circular orbit in the disc
plane has $\epsilon \simeq 1$.  In Fig. \ref{scatter} we show
$\epsilon$ as a function of total energy for the stars in each of the
Aquarius haloes. The colour code indicates stellar metallicity,
[Fe/H].  Discs are easily identified as star concentrations near
$\epsilon = 1$ while stars in the spheroidal components usually
scatter around $\epsilon = 0$. The bulge is located at the lowest
(most bound) energies, while the outer halo is at the highest (least
bound) energies.

We define stars to belong to different dynamical components according
to their location in Fig.\ref{scatter}. We consider stars with
$\epsilon > 0.65$ to be part of a disc.  This value is small enough to
include stars in both thin and thick discs.  Particles which do not
satisfy this requirement are taken to be part of the spheroid.
Motivated by observations of the Milky Way spheroid which report
differences in stellar kinematics and chemical abundances as one moves
outwards \citep{caro2007,zoca2008}, we separate our spheroids into
three subregions according to their binding energy.  The central
spheroid (``bulge'') is defined to consist of stars more bound than
the minimum energy ($E_{\rm cen}$) of stars with $r\ge 0.5 \times
r_{\rm opt}$ where the characteristic radius $r_{\rm opt}$ is defined
to enclose $83\%$ of the baryonic mass of the simulated
galaxy.
Central spheroid stars thus all have orbital
apocentres less than about $0.5 r_{\rm opt}$. Stars more weakly bound
than $E_{\rm cen}$ but with $E<E_{\rm inner}$, the minimum energy of
all stars with $r>2 \times r_{\rm opt}$ are considered to form the
inner halo. Stars with $E > E_{\rm inner}$ are defined to belong to
the outer halo and thus all have orbital pericentres outside about $2
r_{\rm opt}$.  Although there is some arbitrariness in these criteria,
the main features of the components change only slowly with the
reference energies.  Our criteria are chosen so that the definition of
the spheroidal components adapts to the overall size of each
individual galaxy. Fig.~\ref{scatter} indicates the two bounding
energies($E_{\rm cen}$ and $E_{\rm inner}$) for each Aquarius system.

\begin{figure*}
\hspace*{-0.2cm}\resizebox{4.5cm}{!}{\includegraphics{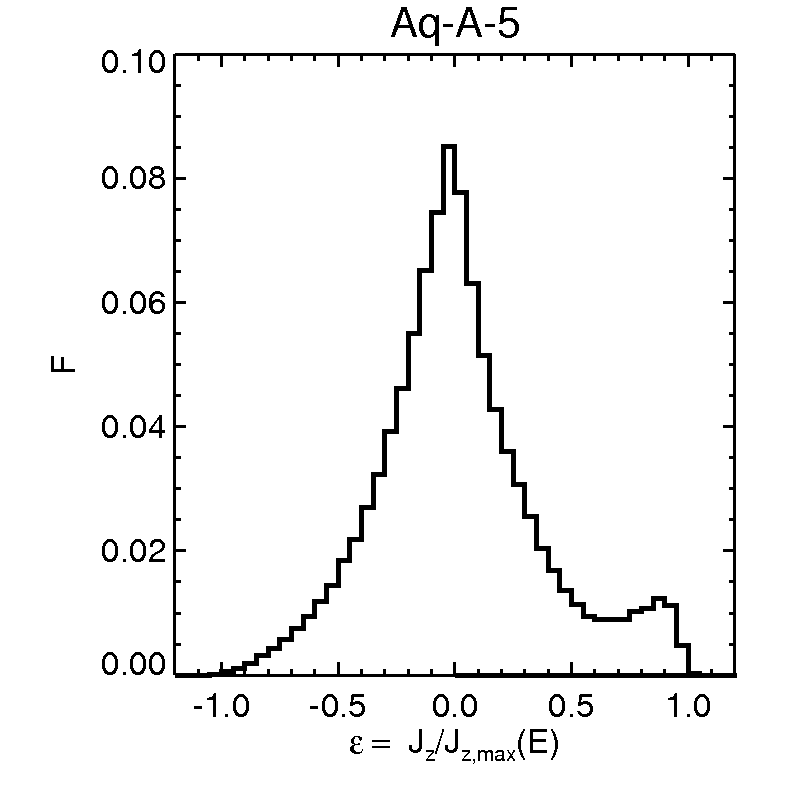}}
\hspace*{-0.3cm}\resizebox{4.5cm}{!}{\includegraphics{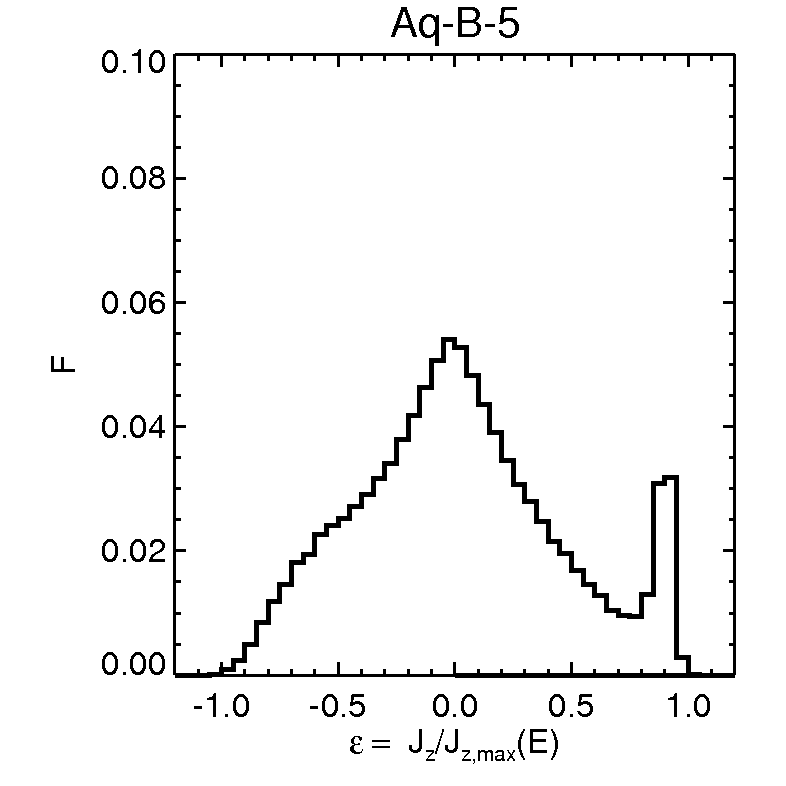}}
\hspace*{-0.3cm}\resizebox{4.5cm}{!}{\includegraphics{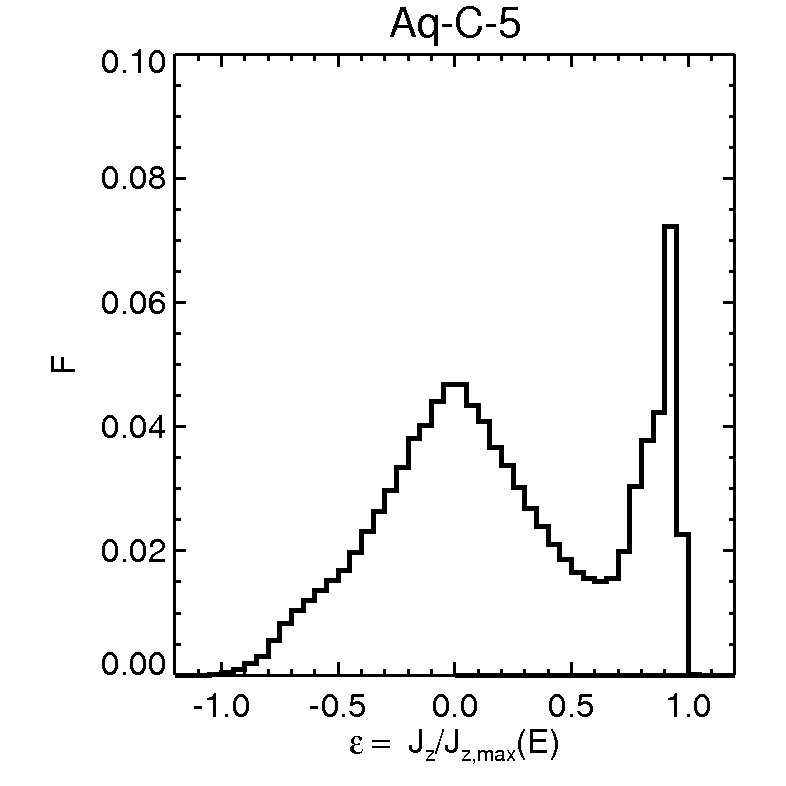}}
\hspace*{-0.3cm}\resizebox{4.5cm}{!}{\includegraphics{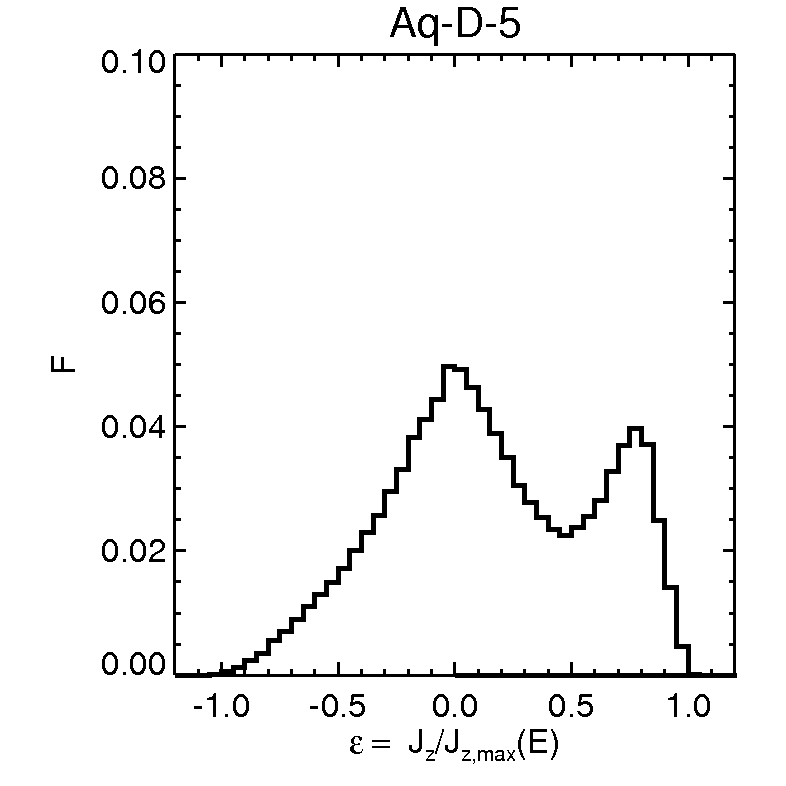}}\\
\hspace*{-0.2cm}\resizebox{4.5cm}{!}{\includegraphics{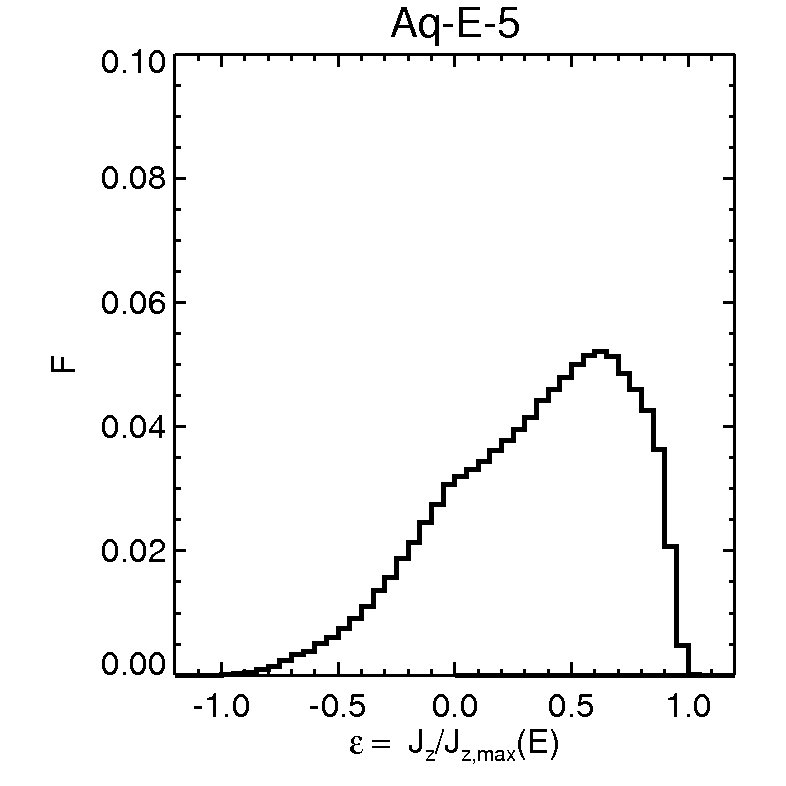}}
\hspace*{-0.3cm}\resizebox{4.5cm}{!}{\includegraphics{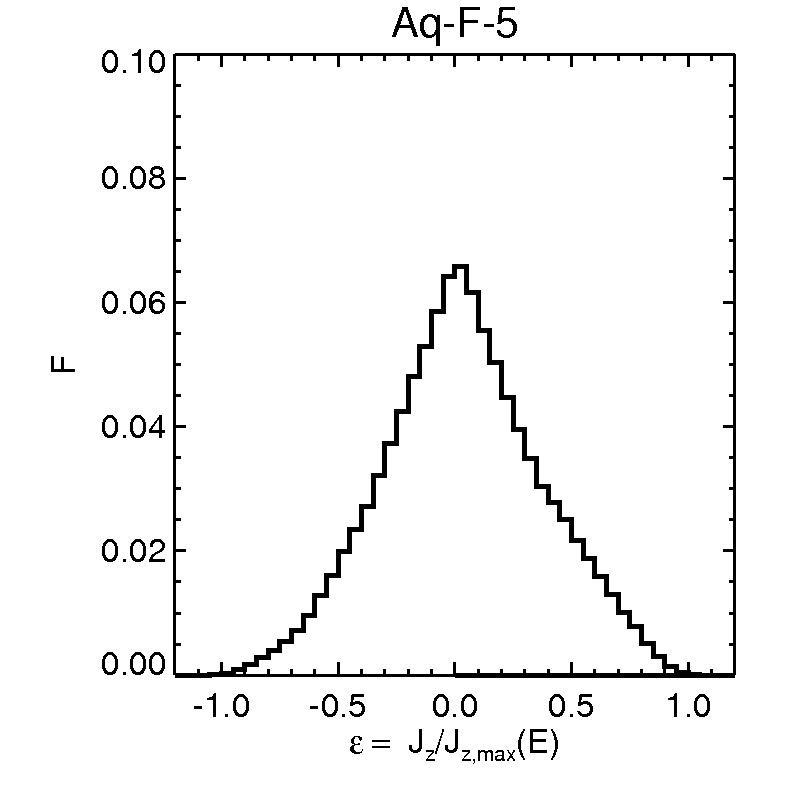}}
\hspace*{-0.3cm}\resizebox{4.5cm}{!}{\includegraphics{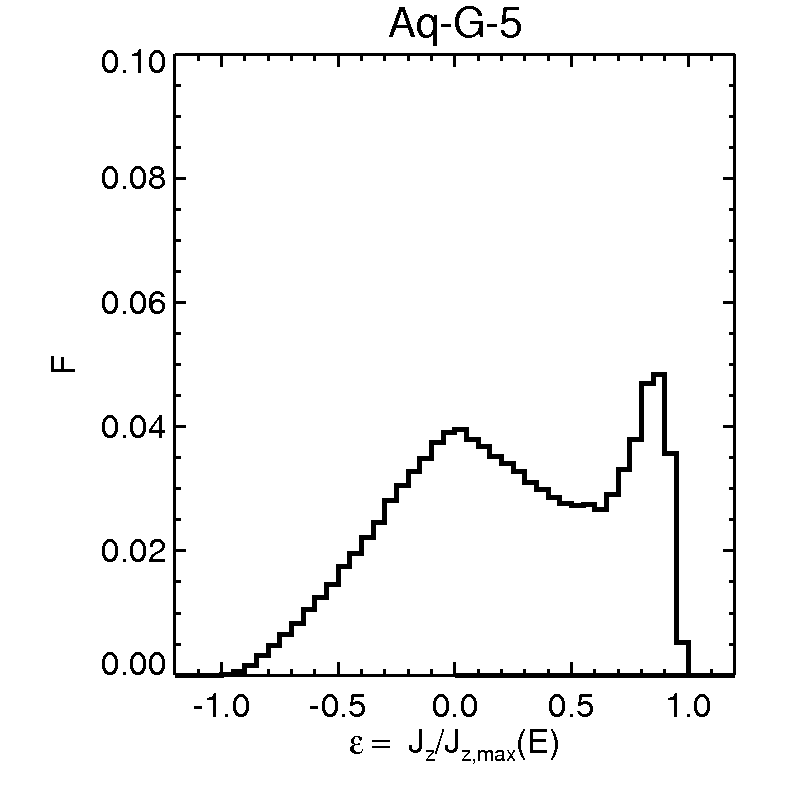}}
\hspace*{-0.3cm}\resizebox{4.5cm}{!}{\includegraphics{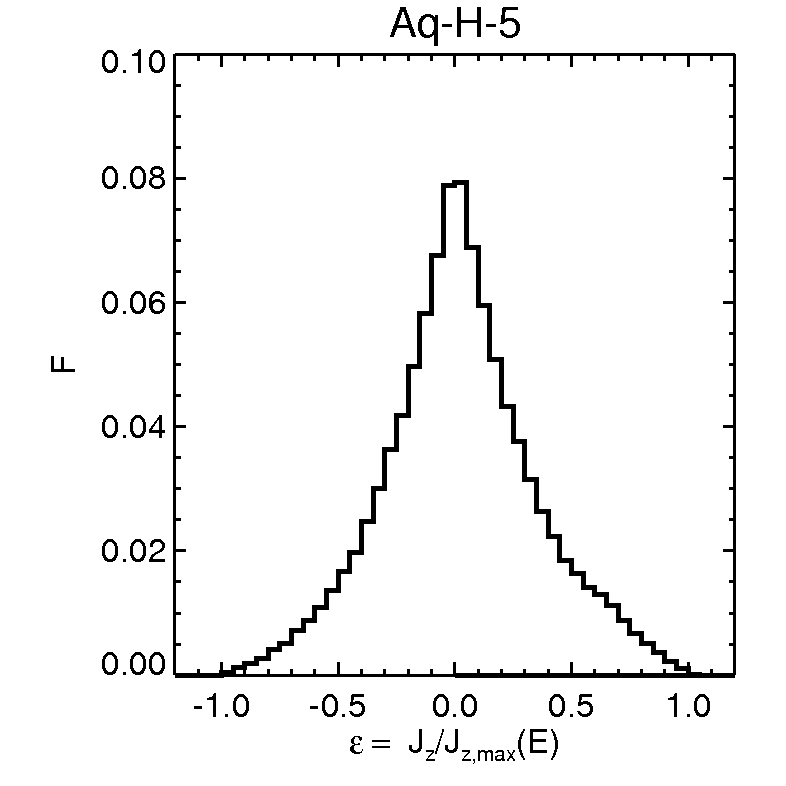}}\\
\caption{Histograms of stellar mass as a function of $\epsilon=J_z/J_{z,
    \rm max}(E)$ for our Aquarius haloes within twice $r_{\rm gal}$ at
  $z=0$. Stars in satellites have been excluded.  }
\label{histJ}
\end{figure*}

The disks extend over a relatively small range of energies and thus
radii in Aq-A and B, but over a more substantial range in Aq-C, D, E
and G. Aq-F and H appear to have very little disc at all. The relative
mass of the discs is seen more easily in Fig.\ref{histJ} where we give
histograms of $\epsilon$ for all stars within $2r_{\rm opt}$ in each
of our galaxies. Returning to Fig\ref{scatter}, the central spheroid
is seen to have rather little rotation in Aq-A, C, D, F, G and H, but
is clearly rotating in the same sense as the disc in Aq-E.  In Aq-B
the inner spheroid is rotating in the same sense as the disc in the
innermost regions but in the opposite sense at intermediate energies.

The inner stellar halo is essentially non-rotating in all cases except
Aq-E and F. In the former there is no clear separation of the inner
halo and disk populations, as is particularly evident in
Fig.\ref{histJ}. The outer halo is non-rotating in all cases except
Aq-F, and in most galaxiees it shows stellar concentrations associated
with satellite galaxies and stellar streams (i.e. disrupted
satellites). For our main analysis we remove all satellites detected
as individual bound structures in order to focus on the ``diffuse''
stellar halo.  The chemical abundances of surviving satellites are
analysed separately in the last section of this paper. Colour
variations between components in different panels of Fig.\ref{scatter}
indicate different levels of chemical enrichment. In general, strongly
bound stellar populations are more chemically enriched than those
found at lower energies. However, disc stars are clearly more
chemically enriched than spheroid stars of similar energy.

\section{Formation histories of the stellar components and their global chemical properties}

It has long been suggested that particular chemical patterns are linked to key
events in the formation history of galaxies.  In this section, we will address
this question by comparing formation history with $z=0$ abundance pattern for
the various components of our galaxies.

\begin{figure*}
\hspace*{-0.2cm}\resizebox{4.5cm}{!}{\includegraphics{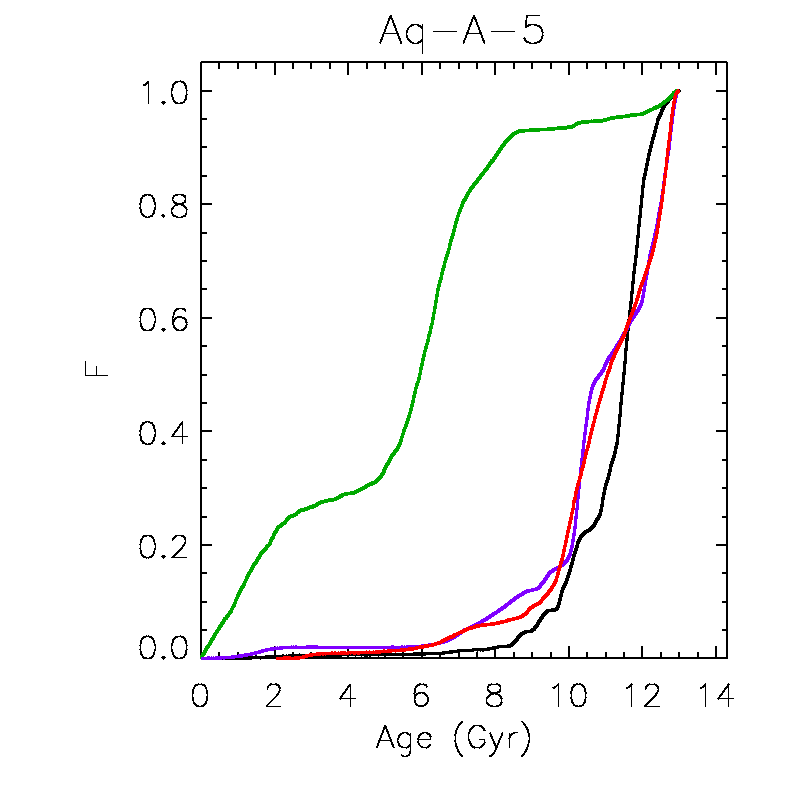}}
\hspace*{-0.3cm}\resizebox{4.5cm}{!}{\includegraphics{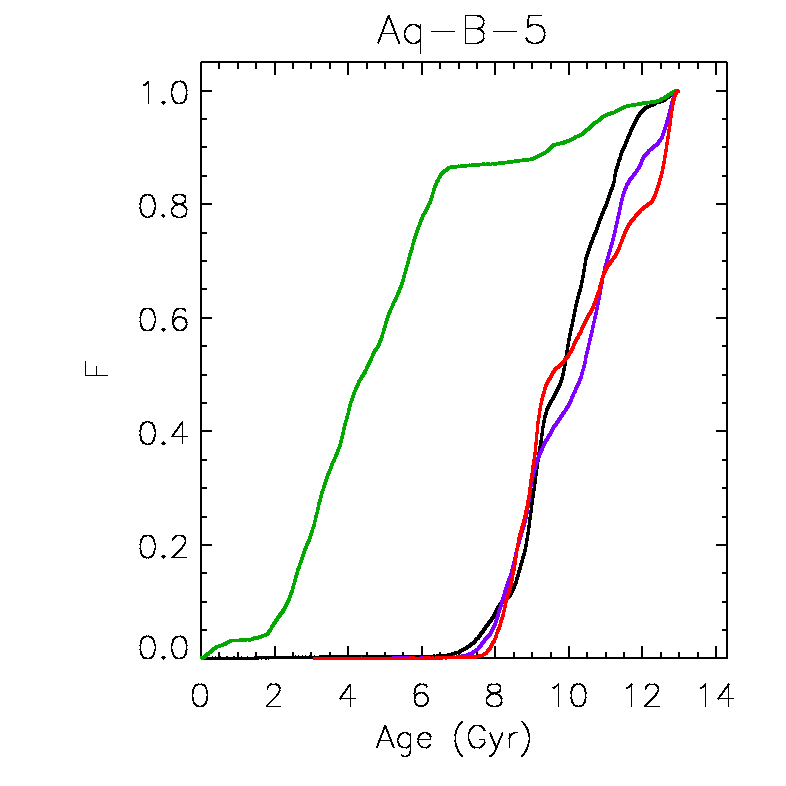}}
\hspace*{-0.3cm}\resizebox{4.5cm}{!}{\includegraphics{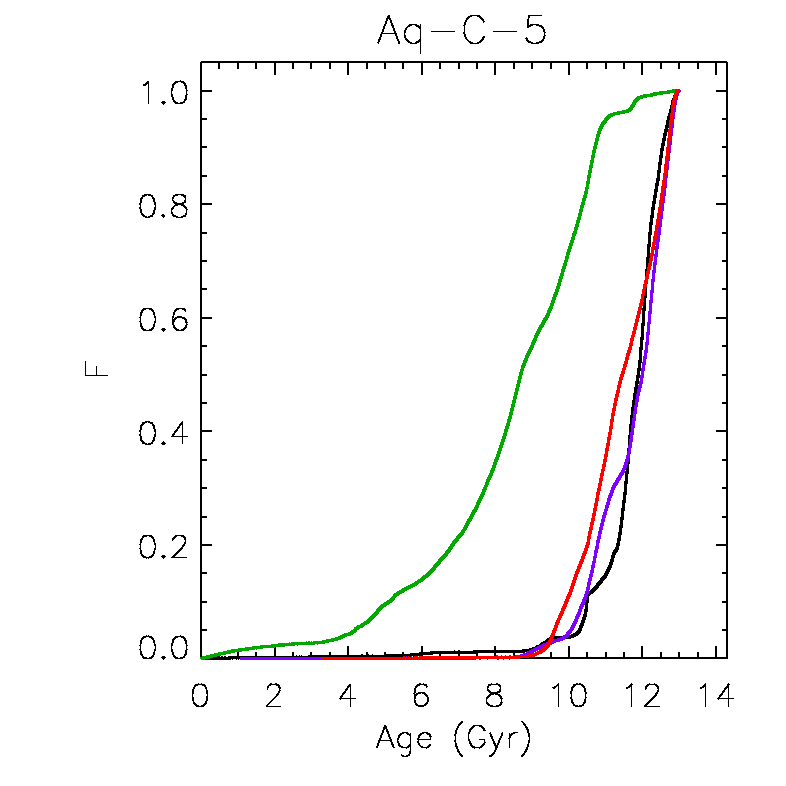}}
\hspace*{-0.3cm}\resizebox{4.5cm}{!}{\includegraphics{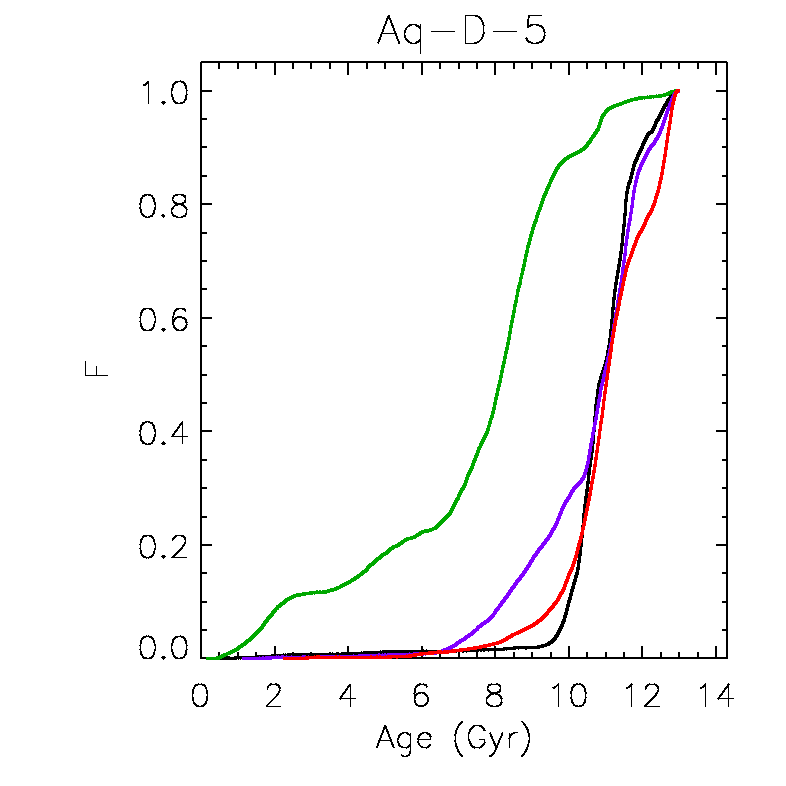}}\\
\hspace*{-0.2cm}\resizebox{4.5cm}{!}{\includegraphics{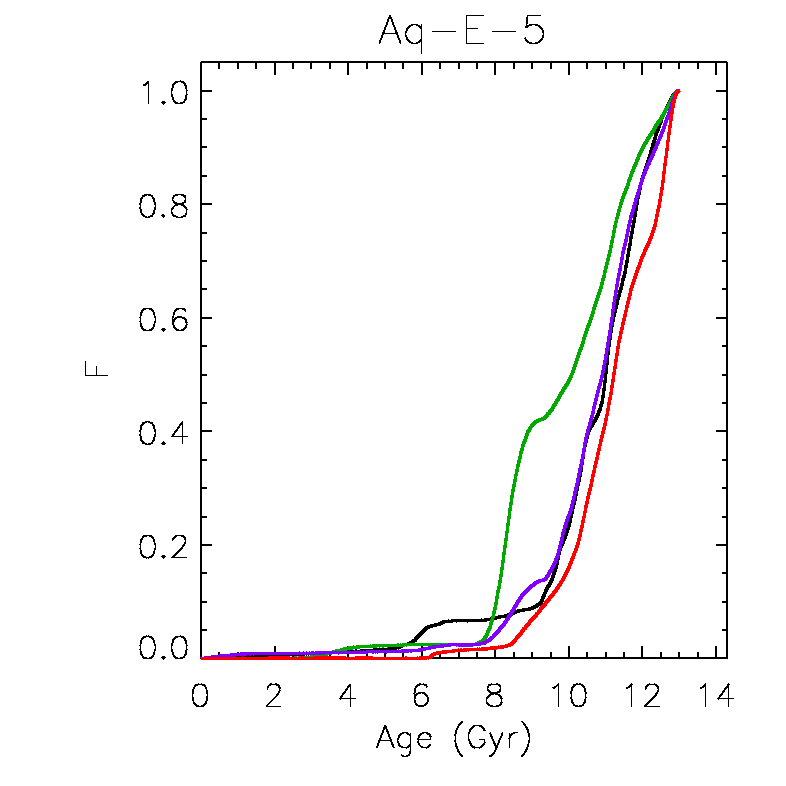}}
\hspace*{-0.3cm}\resizebox{4.5cm}{!}{\includegraphics{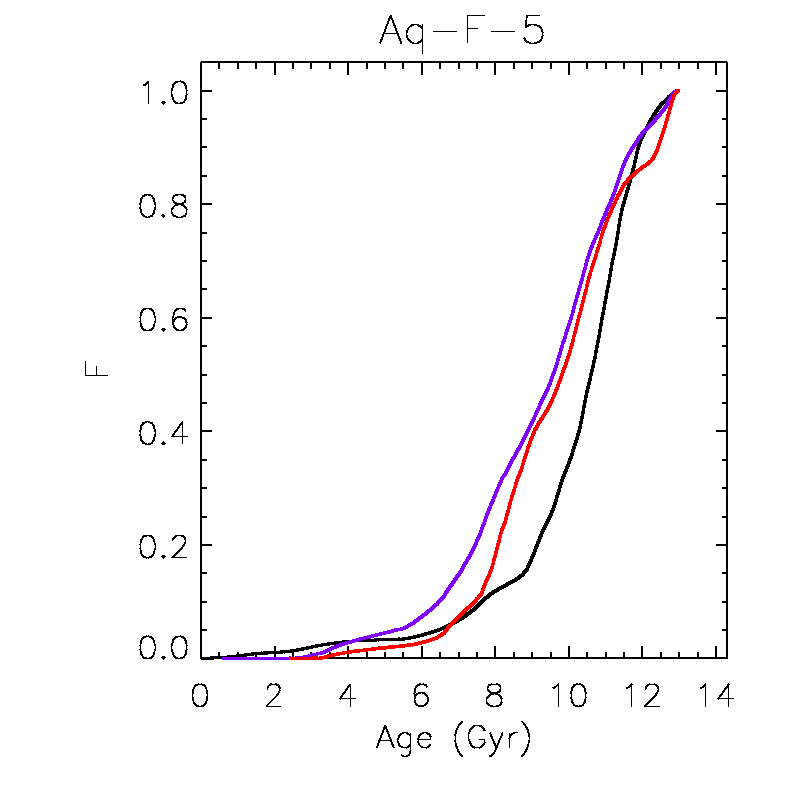}}
\hspace*{-0.3cm}\resizebox{4.5cm}{!}{\includegraphics{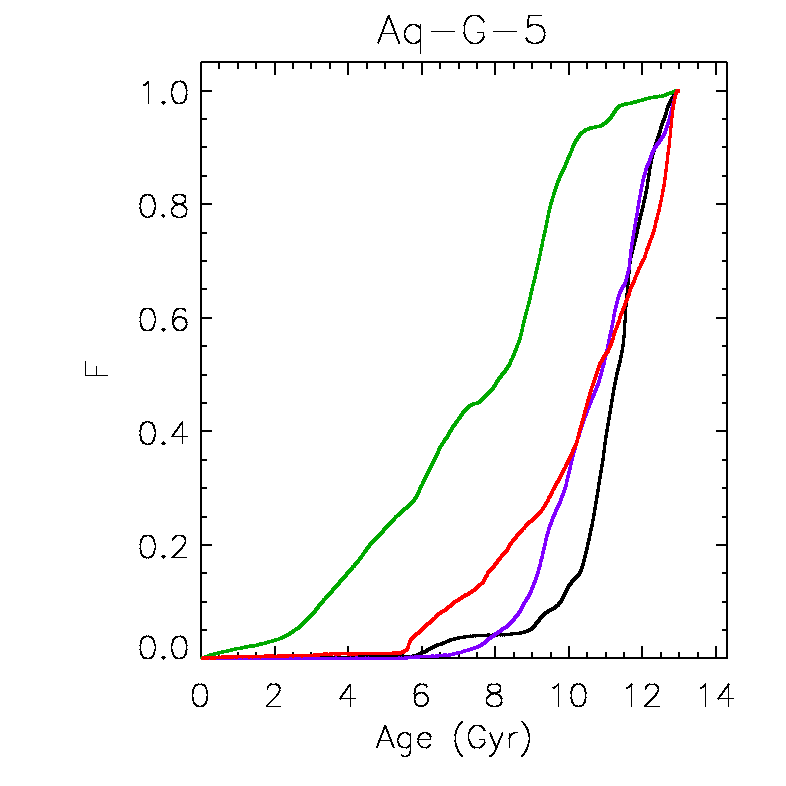}}
\hspace*{-0.3cm}\resizebox{4.5cm}{!}{\includegraphics{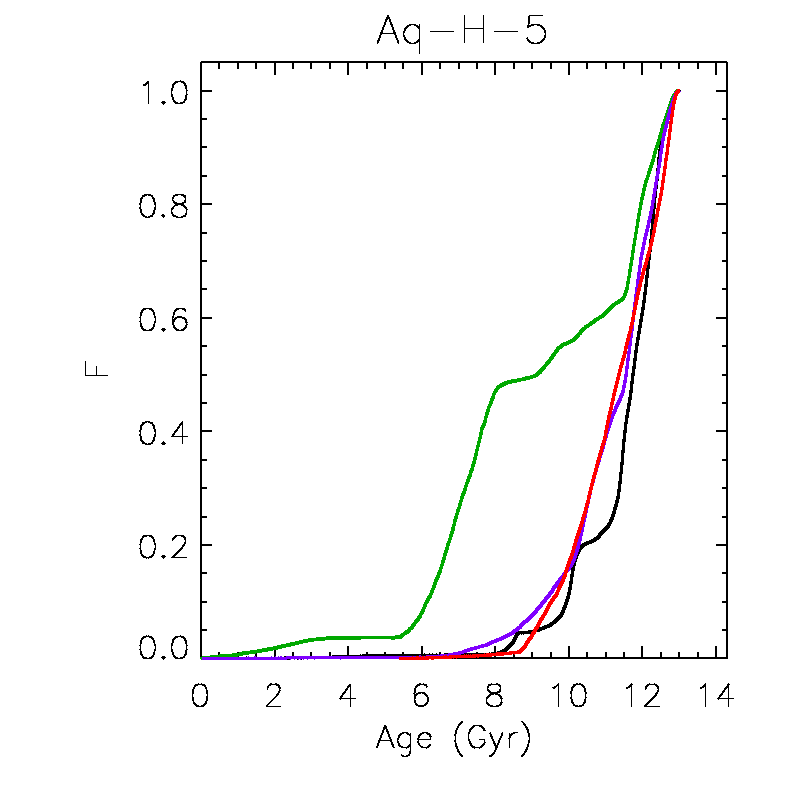}}
\hspace*{-0.2cm}
\caption{Cumulative stellar mass as a function of age for stars in the
  disc (green), the central spheroid (black), the inner halo (violet)
  and the outer halo (red) for all simulated galaxies. }
\label{sfrhistories}
\end{figure*}

In Fig. ~\ref{sfrhistories}, we present star formation histories for
individual stellar components in the form of cumulative plots of the
fraction of the final stellar mass formed as a function of age.  In
such plots, a step reflects a starburst.  The central spheroid and the
inner and outer haloes have quite similar SF histories, with most of
their stars formed at very high redshift.  The oldest stars are split
between the central spheroid and the haloes.  The discs contain
younger stellar populations which form in a series of episodes spread
over the Hubble time. In the case of Aq-F-5, no disc component
survives at $z=0$.

\begin{figure*}
\hspace*{-0.2cm}\resizebox{4.5cm}{!}{\includegraphics{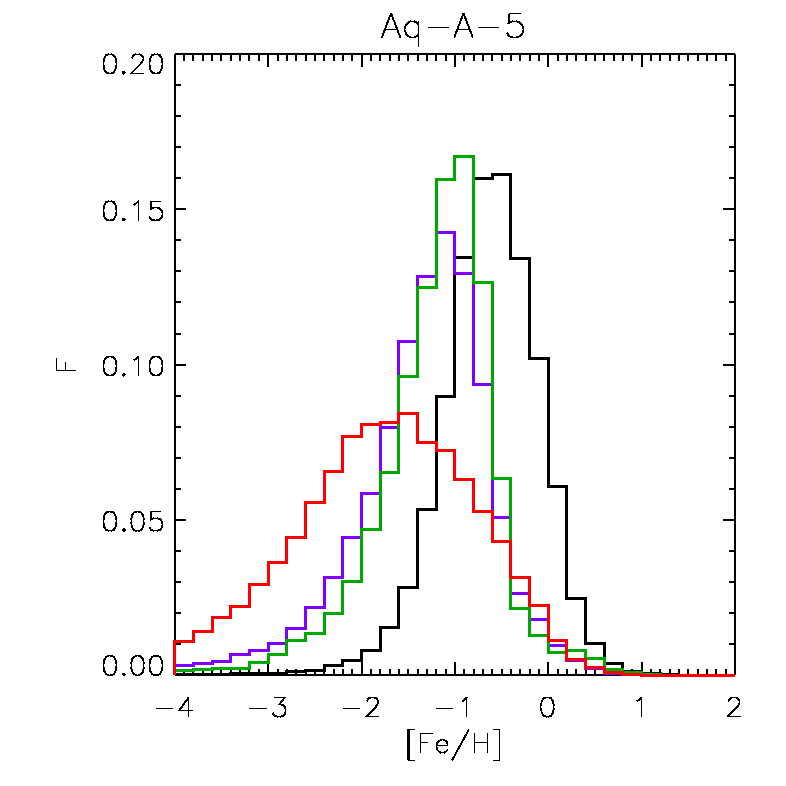}}
\hspace*{-0.2cm}\resizebox{4.5cm}{!}{\includegraphics{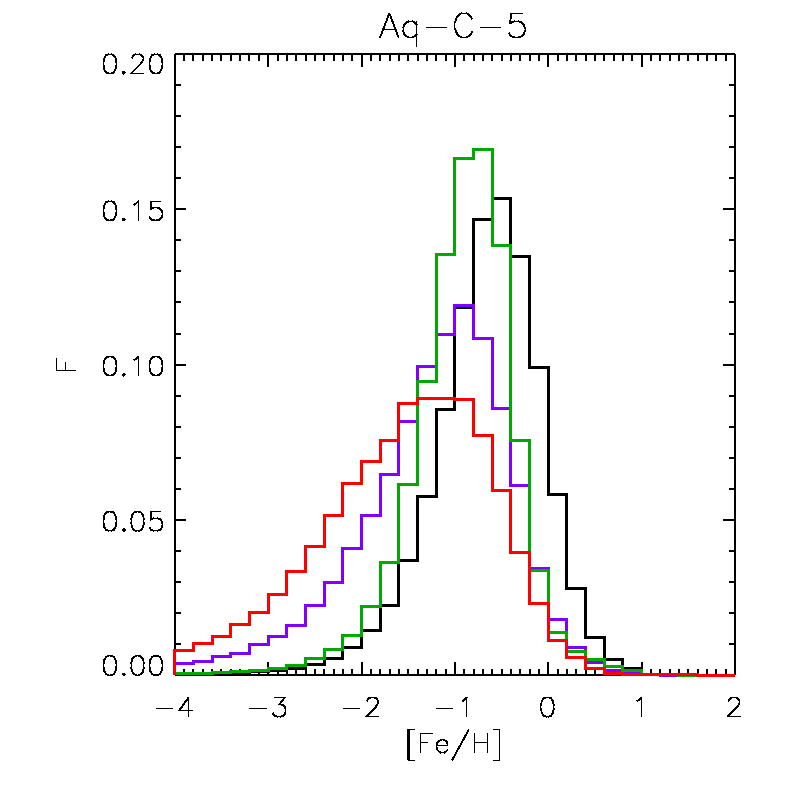}}
\hspace*{-0.2cm}\resizebox{4.5cm}{!}{\includegraphics{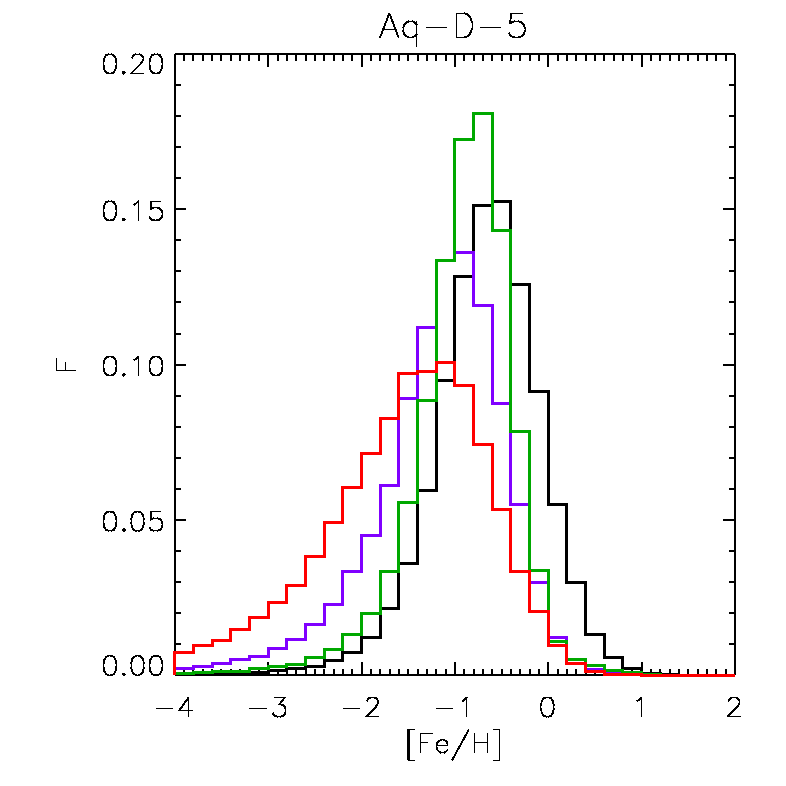}}
\hspace*{-0.2cm}\resizebox{4.5cm}{!}{\includegraphics{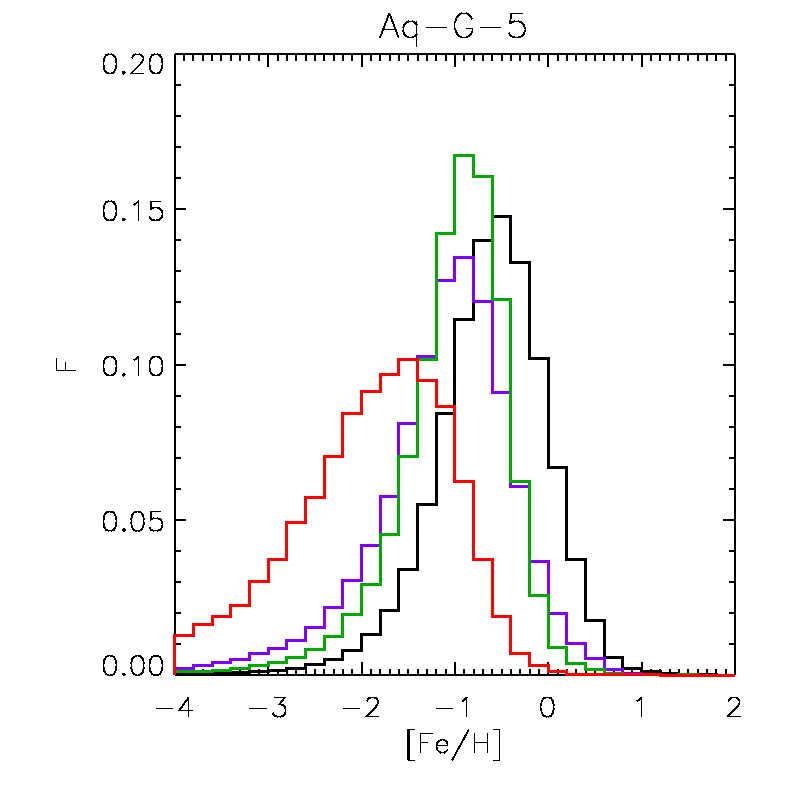}}\\
\hspace*{-0.3cm}\resizebox{4.5cm}{!}{\includegraphics{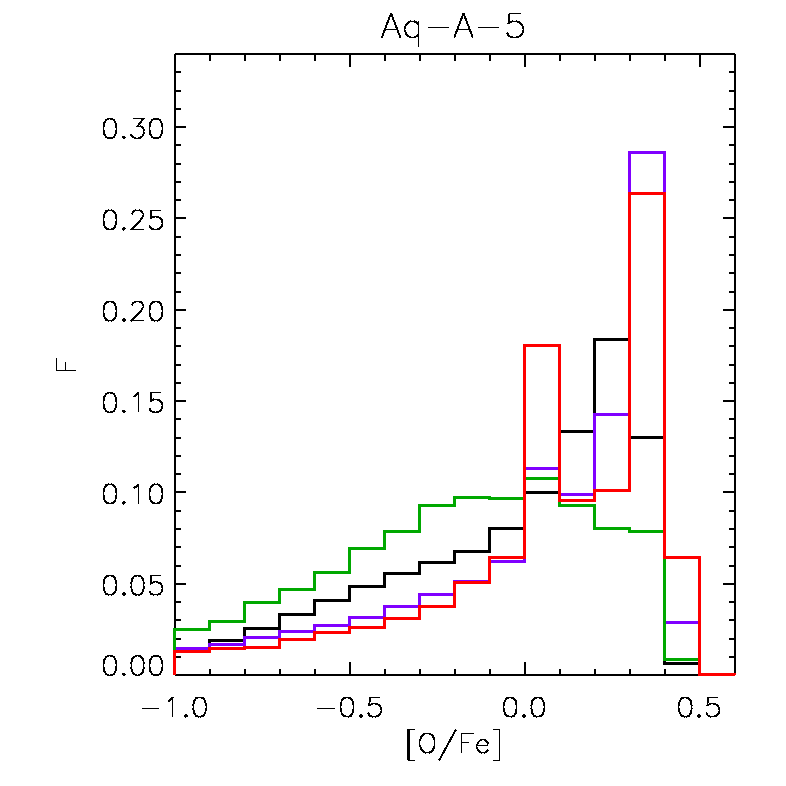}}
\hspace*{-0.3cm}\resizebox{4.5cm}{!}{\includegraphics{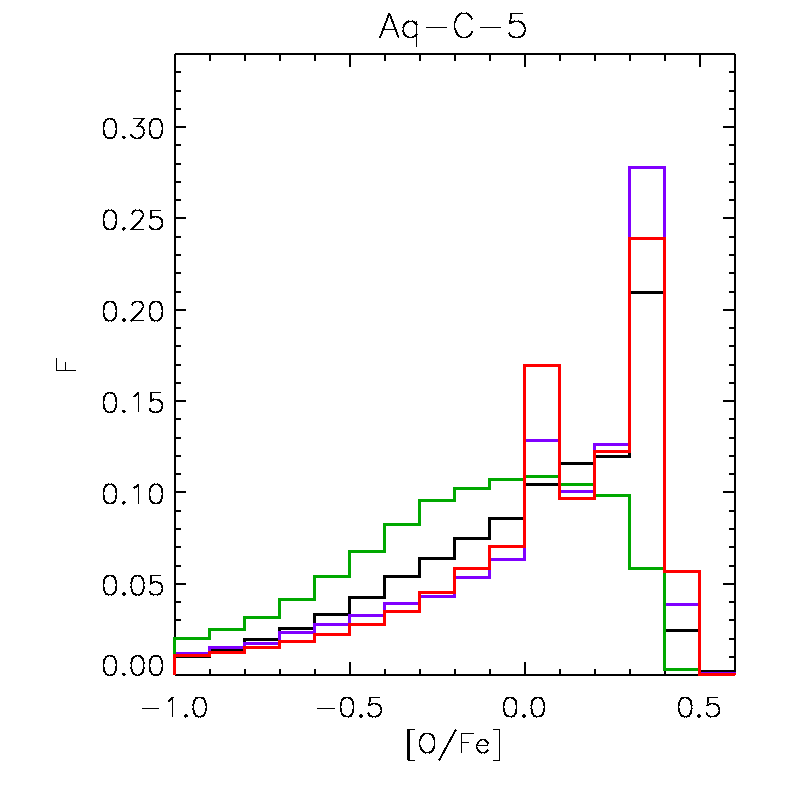}}
\hspace*{-0.3cm}\resizebox{4.5cm}{!}{\includegraphics{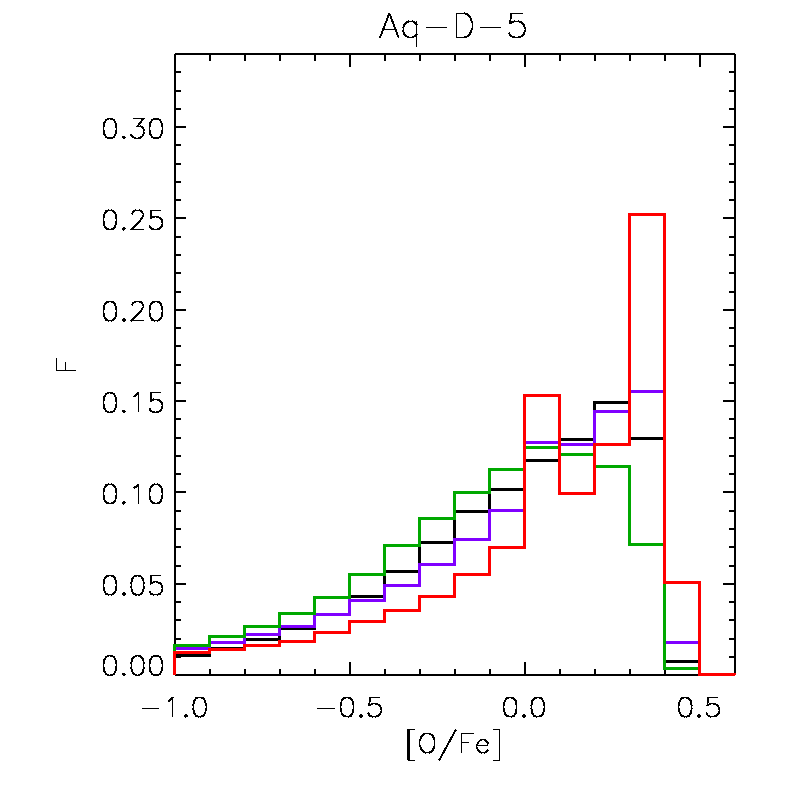}}
\hspace*{-0.3cm}\resizebox{4.5cm}{!}{\includegraphics{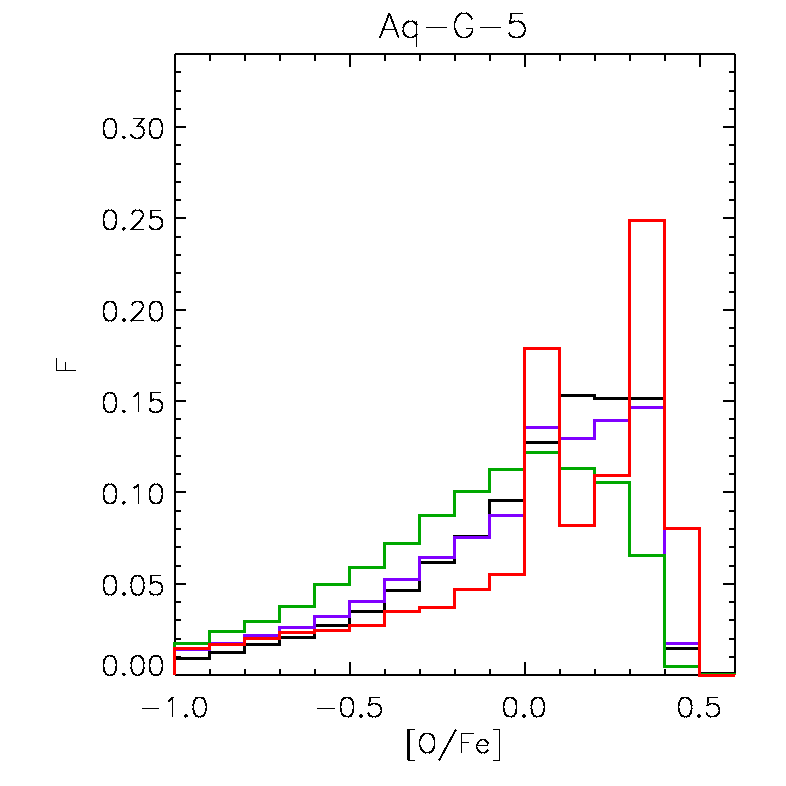}}\\
\hspace*{-0.2cm}
\caption{ Distributions of [Fe/H] (upper raw) and [O/Fe] (lower
  raw) for stars in the discs (green), central spheroids
  (black),inner stellar haloes (violet) and outer stellar haloes (red)
  of four Aquarius galaxies. }
\label{fehtot}
\end{figure*}

Star formation histories vary from system to system as a consequence
of differing evolutionary paths, leading to variations in chemical
properties between our galaxies and between components within a
galaxy.  This can be seen in Fig. ~\ref{fehtot} where we show [Fe/H]
distributions for the individual stellar components in four of our
simulated haloes.  Interestingly, while the central spheroids, the
inner and the outer stellar haloes all have similar star formation
histories (see Fig. ~\ref{sfrhistories}) their [Fe/H] distributions
are very different.  Comparing individual stellar components between
galaxies, we also note significant variations in the abundance
distributions.

The relative abundance of $\alpha$-elements with respect to Fe is
often identified as a key property to characterise the formation
histories of stellar populations.  We have measured [O/Fe] distributions
for the individual stellar components in four of our galaxies as displayed
in Fig. ~\ref{fehtot} (upper raw). As in the Milky Way, discs show the
lowest level of $\alpha$-enhancement. This is because their stars
are younger than the other simulated stellar populations and 
formed over an extended period of time.
As a result, the gas from which they formed had time to be enriched by
SNIa.  Stars in the central spheroids are $\alpha$-enhanced but have
the highest [Fe/H] values of all the populations, reflecting the fact
that they formed in a short and strong starburst and were able to
retain a large fraction of their SN ejecta. These general properties
are in agreement with data for the Milky-Way bulge \citep{zoca2008}.
The largest $\alpha$-enhancements in our models are, however, found in
the inner and outer stellar haloes.

Although the typical [Fe/H] values found for
our bulge and disc components are lower than for their Milky Way counterparts,
the trends between components are  similar to those observed.  
There are
clear variations between the various Aquarius galaxies, both in [Fe/H] and in
$\alpha$-enhancement, but the above mentioned trends seem to be common to all of them.
Remaining differences with the Milky Way  (for example the G-dwarf problem evident  in Fig. ~\ref{fehtot})
must, at least in part, be due to the
fact that all the Aquarius galaxies have disc-to-spheroid mass ratios much
smaller than that of the Milky Way.
 We also note that the metallicity distributions have large standard deviations which might suggest the
need for  more efficient metal mixing \citep{mart2008,wier2009}.
For the discs and inner haloes, these large dispersions 
 make it  difficult to distinguish  chemical differences. 
 Nevertheless, for the best discs, Aq-C-5,
Aq-D-5 and Aq-G-5,  a Kolmogorov-Smirnov test shows the
 metallicity distributions  of the discs and inner haloes to differ at confidence levels between 70 and 94 percent.
 Only for Aq-B-5 and Aq-H-5  do the the metallicity distributions of these
components appear indistinguishable. 
In general,  for all our galaxies, the spheroidal components show a decrease of [Fe/H] 
as one moves outwards. Their stellar populations are more $\alpha$-enriched and older
than those populating the corresponding disc components. 
Stars in the haloes are, on average, more $\alpha$-enriched
than those in the central spheroids. 

To test how numerical resolution
affects our estimates of chemical abundances, we have analysed two
lower resolution versions of Aq-E which we label Aq-E-6 and
Aq-E-7 \footnote{Simulations Aq-E-6 and Aq-E-7 correspond to those labeled as Aq-E-6b and Aq-E-6 in \citet{scan11}}. Within $r_{200}$ these have totals of about 160,000 and 80,000
particles, respectively, whereas Aq-E-5 has more than one
million. These two lower resolution simulations were also analysed
by. \citet{scan09,scan11} and \citet{tissera2010}
who reported good agreement between their dynamical and
structural properties and those of Aq-E-5, both for baryons and for
dark matter. We find similarly good convergence for the chemical
properties of each of our stellar components, as can be appreciated
from Fig.5 where we compare [Fe/H] distributions in the three
simulations.

\begin{figure*}
\hspace*{-0.2cm}\resizebox{4.5cm}{!}{\includegraphics{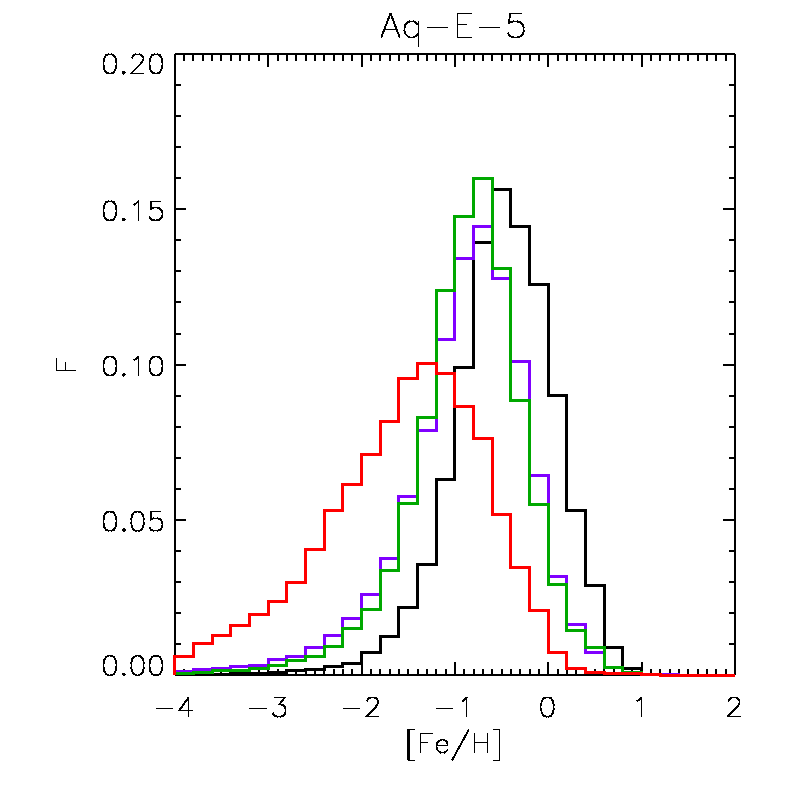}}
\hspace*{-0.2cm}\resizebox{4.5cm}{!}{\includegraphics{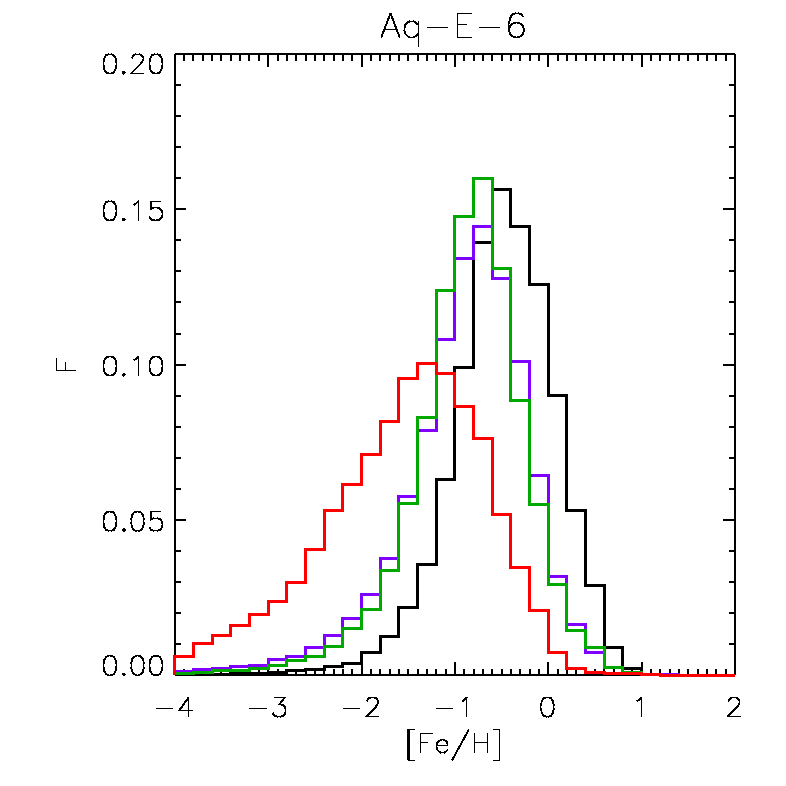}}
\hspace*{-0.2cm}\resizebox{4.5cm}{!}{\includegraphics{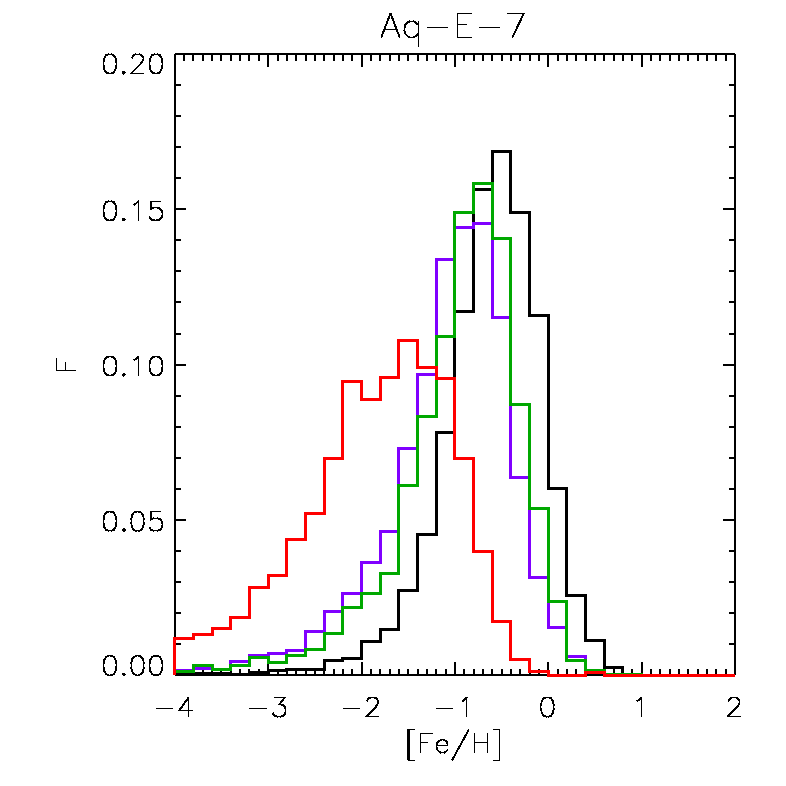}}
\hspace*{-0.2cm}
\caption{ Distributions of [Fe/H]  for stars in the discs (green), central spheroids
  (black),inner stellar haloes (violet) and outer stellar haloes (red)
  of Aq-E sampled at three different levels of resolution, declining from Aq-E-5
to Aq-E-7. }
\label{resol}
\end{figure*}
 
\subsection{Assembly histories}

In order to understand if the stars in each component were formed {\it in
  situ} (i.e.  in the main progenitor of the final galaxy) or in smaller
objects which were subsequently accreted, we follow each star from its time of
formation to $z=0$.  This allows us to estimate the fraction of stellar mass
in each component which formed {\it in situ}.  We consider all self-bound
bound clumps of more than 20 simulation particles to be separate objects, so
that any star which formed in a clump other than the main progenitor is
considered to have been accreted.

\begin{figure*}
\resizebox{7cm}{!}{\includegraphics{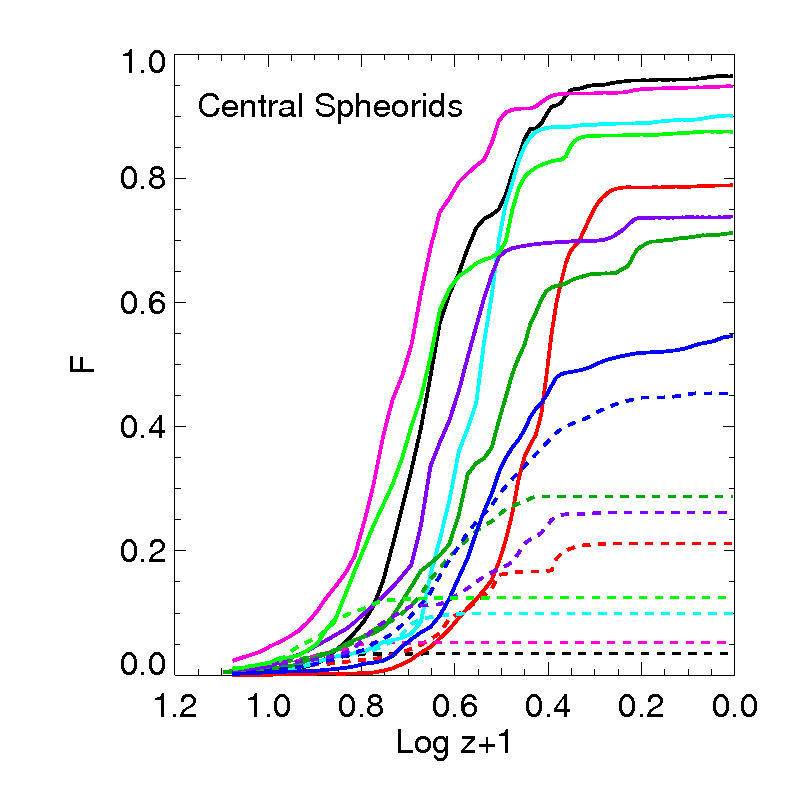}}
\resizebox{7cm}{!}{\includegraphics{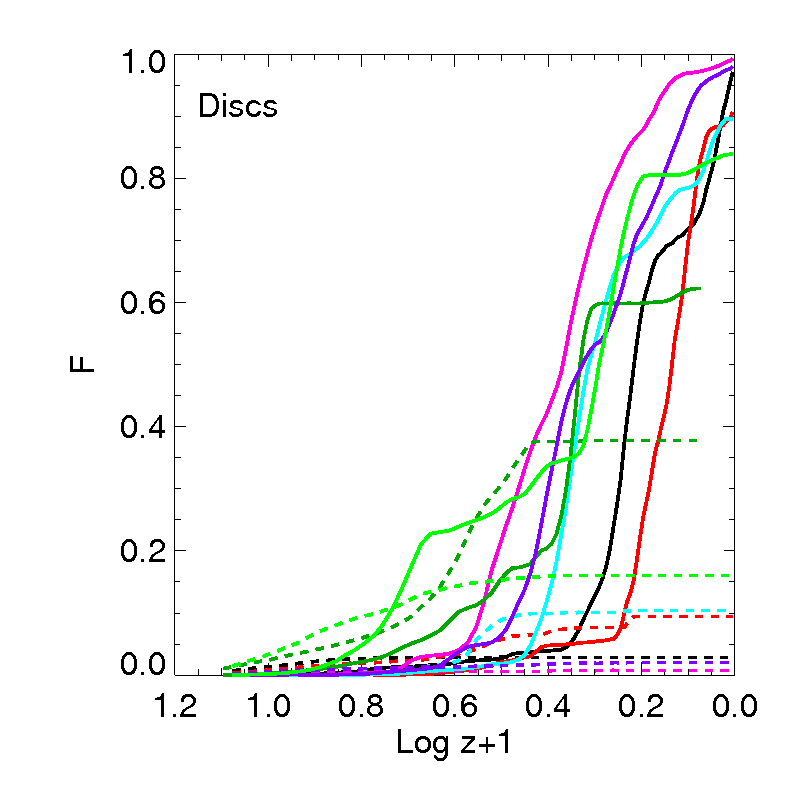}}\\
\resizebox{7cm}{!}{\includegraphics{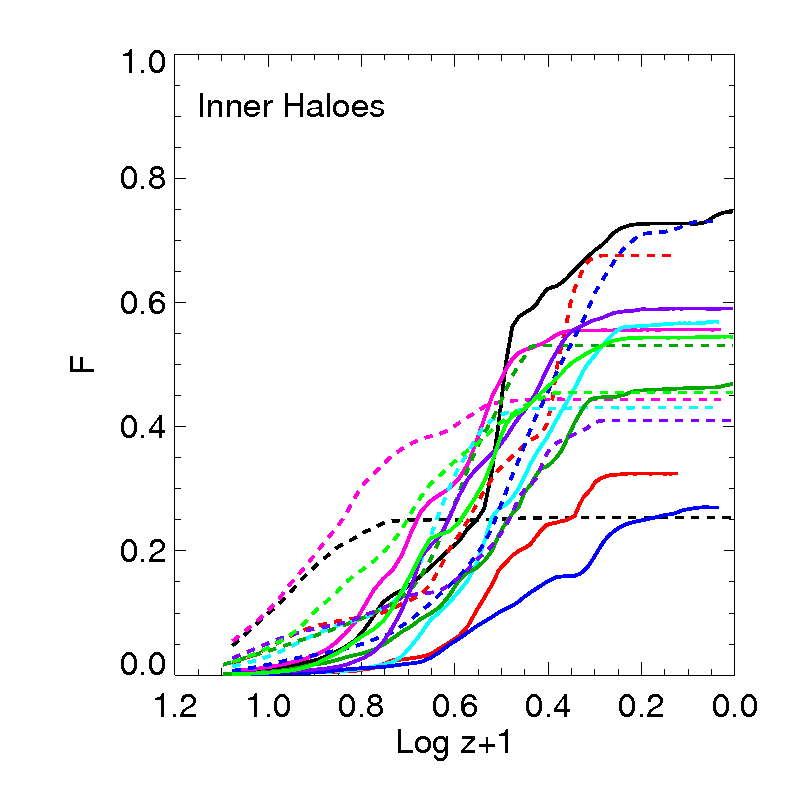}}
\resizebox{7cm}{!}{\includegraphics{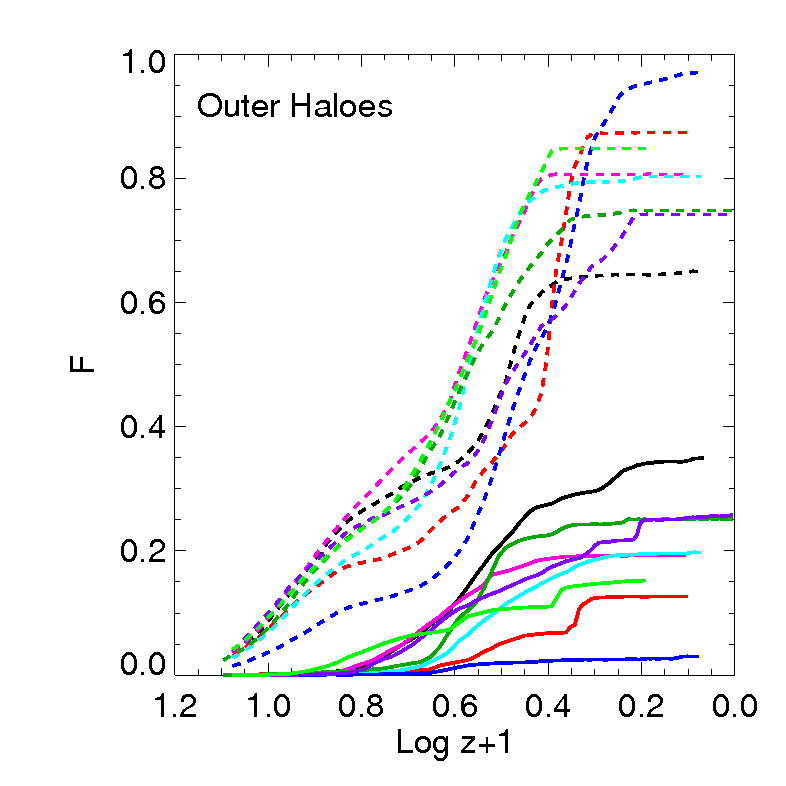}}
\hspace*{-0.2cm}
\caption{Accreted (dashed) and {\it in situ} (solid) fractions as a
  function of formation redshift for stars in each of our four
  components and for all eight of the Aquarius galaxies. Each of the
  panels corresponds to a different component as labelled, while the
  colours of the curves identify the individual galaxies. The key is:
  Aq-A-5 (black), Aq-B-5 (red), Aq-C-5 (magenta), Aq-D-5 (cyan),
  Aq-E-5 (green), Aq-F-5 (blue), Aq-G-5 (violet) and Aq-H-5 (light
  green).  }
\label{fracprog}
\end{figure*}

Cumulative {\it in situ} and accreted fractions are shown in
Fig. ~\ref{fracprog} as a function of the redshift at which the stars
formed. The four panels give results for our four stellar components,
and within each panel solid lines give {\it in situ} fractions while
dashed lines give accreted fractions. Different colours refer to the
different Aquarius galaxies as noted in the caption. Most of the
stellar mass in the central spheroids was formed {\it in situ} varying
from $\sim 60$ to $\sim 95$ per cent. Only in Aq-F-5, which underwent
a major merger at $z\sim 0.6$, are the {\it in situ} and accreted
fractions comparable. In contrast, in the outer stellar haloes the
majority of the stars are accreted in all eight galaxies. (Remember
that we have excluded stars which remain part of a bound satellite at
$z=0$ when compiling these statistics.)  The inner haloes show an
intermediate behaviour with both fractions ranging between 25 and 75
per cent. About 75 per cent of the inner halo of Aq-F-5 is accreted,
most of it as part of the second object in the major merger.  The
$z=0$ discs are heavily dominated by {\it in situ} stars, with a
maximum accreted contribution of $\sim 15$ per cent. Note however,
that the largest accreted fractions are for Aq-E-5, where Figs
\ref{scatter} and \ref{histJ} show there to be no clear separation
between disc and inner halo, and Aq-H-5, where the disc component is
very weak.

\begin{figure}
\hspace*{1cm}\resizebox{7cm}{!}{\includegraphics{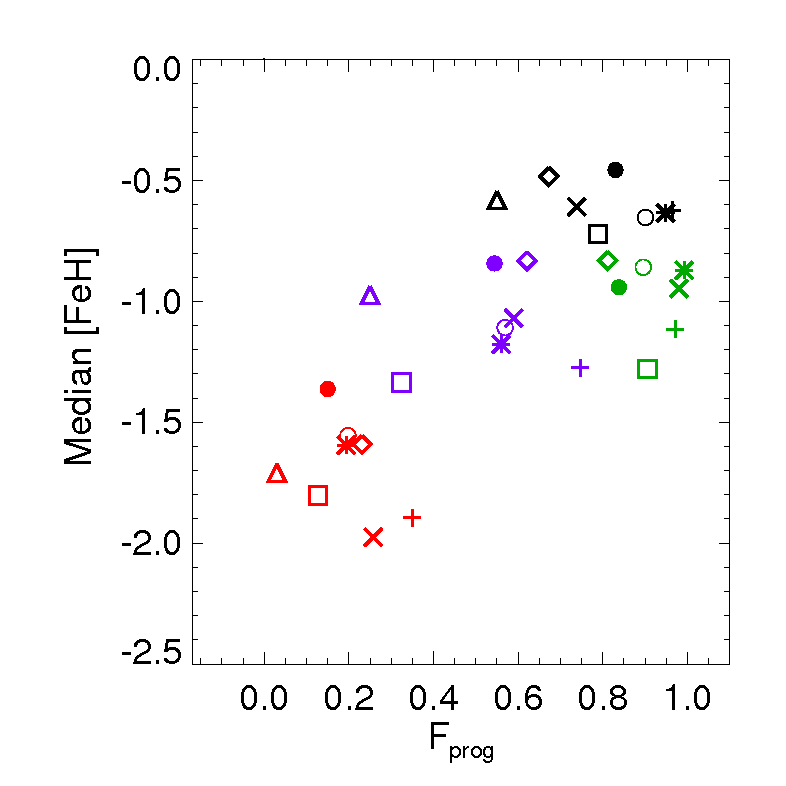}}
\hspace*{1cm}\resizebox{7cm}{!}{\includegraphics{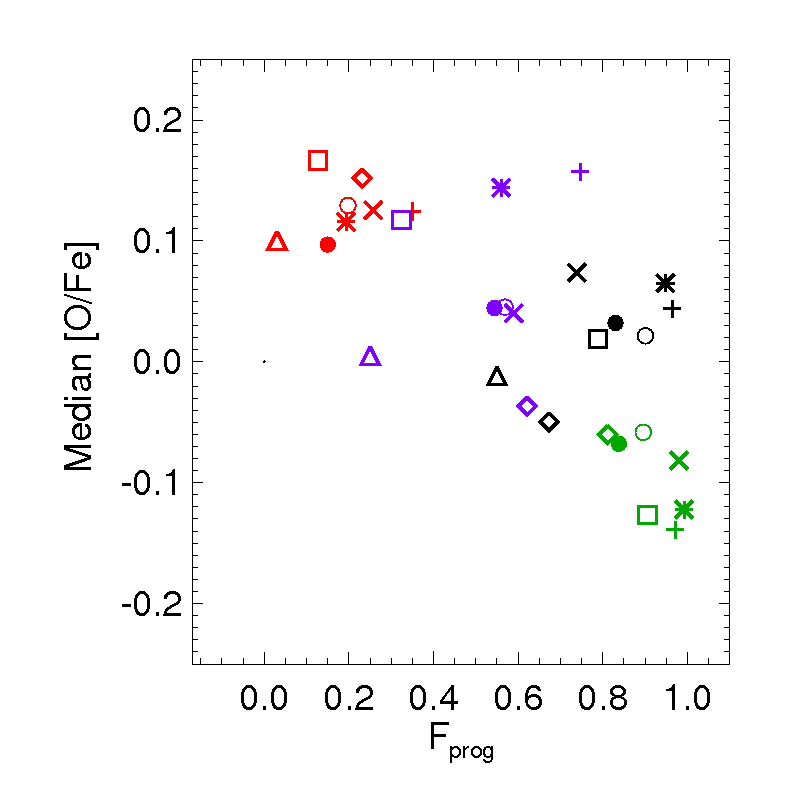}}
\hspace*{-0.2cm}
\caption{ Medians of [Fe/H] and [O/Fe] as a function of $F_{\rm prog}$
  for central spheroids (black), discs (green), inner haloes (violet)
  and outer halos (red). Symbols differentiate the eight galaxies:
  Aq-A-5 (cross), Aq-B-5 (square), Aq-C-5 (asterisk), Aq-D-5 (open
  circle), Aq-E-5 (diamond), Aq-F-5 (triangle), Aq-G-5 (ex) and
  Aq-H-5 (filled circle).}
\label{medianas}
\end{figure}

To correlate assembly histories with chemical abundances, we compare the
fraction of stars formed {\it in situ} ($F_{\rm prog}$) in each stellar
component to the median abundance of the stars at $z=0$. In
Fig. ~\ref{medianas} we show medians of [Fe/H] as a function of $F_{\rm
  prog}$ (upper panel).  As can be seen, stellar components become
more metal-rich as one moves inward; the central spheroid is always much
more chemically enriched than the corresponding outer halo. The trends are
encouragingly consistent with Milky Way data.  The median [Fe/H] of central
spheroids appears independent of {\it in situ} fraction, whereas in the
outer halo is there a clear trend for higher [Fe/H] to correspond to smaller
{\it in situ} fraction.

In the lower panel of Fig. ~\ref{medianas} we plot median [O/Fe] against ${\rm
  F_{\rm prog}} $.  The weak $\alpha$-enhancement of disc stars is a
consequence of the bursty and extended star formation activity which gives
rise to this component, allowing new stellar populations to form from material
enriched by SNIa.  Our SN feedback model allows us to track gas particles
which have been promoted from the cold to the hot phase, thus becoming part of
a galactic wind.  We have checked how much of each disc formed
from gas that was ever part of a wind, finding this to vary between $\sim 10$  and $\sim 30$ per cent. Hence, the great majority of disc stars formed from
material that was acquired directly by infall and was enriched {\it in situ}
without cycling through a galactic fountain.

In central spheroids the level of $\alpha$-enrichment increases with {\it in
  situ} fraction as indicated by the median abundances in Fig. ~\ref{medianas}
(lower panel).  This is also consistent with the age distribution of these
stars which, in general, are old, forming in strong starbursts which are too
short for SNIa enrichment to be important (Fig. ~\ref{sfrhistories}).  We find
that more than 40 per cent of the stars in our central spheroids formed from
gas that had been promoted at least once, showing that these dense and
strongly bound regions are able to retain much of their SN ejecta. Since these
regions are $\alpha$-enhanced, the cooling and re-accretion of the promoted
gas particles had to occur in less than $\approx 1$ Gyr.

The outer and inner stellar haloes are more $\alpha$-enhanced than the central
spheroids.  For these components, the fraction of stars formed from gas that
had participated in a wind varies from 5 to 50 per cent, with the largest
fractions of such stars in the inner haloes. We found that these
'recycled-material' stars are primarily accreted rather than {\it in situ}
objects, so were typically formed in satellite galaxies which were later
disrupted.  This is a consequence of our SN feedback model which results in
larger promotion rates (relative to the star formation rate) in systems with
shallower potential wells: smaller haloes experience more significant
winds \citep{scan08,sawala2010,dero10,sawala2011}.

So far we have analysed the abundance distribution of our stellar
components without regard to where the stars formed.  However, there
are different levels of $\alpha$-enrichment in the three spheroidal
components, even though all are dominated by old populations with
quite similar star formation histories.  This suggests that the
assembly history of each component plays an important role.
Hereafter, we focus on this point.

Properties of the {\it in situ } and accreted subpopulations of each component
in each Aquarius galaxy are listed in Tables ~\ref{tabdisc} to
~\ref{tabhaloex}.  Each table refers to a different component and gives for
each galaxy the stellar fractions formed {\it in situ} and from previously
promoted gas, as well as the median [Fe/H] and [O/Fe] and the mean stellar age
separated into {\it in situ} and accreted components.

The {\it in situ} stars in discs have younger and more diverse ages
than those of any other (sub)population, as can be seen from the
penultimate column in Table \ref{tabdisc}. In particular, the few disc
stars acquired from disrupted satellites are not only older but
formed, on average, over shorter time intervals. As a consequence,
these stars tend to have lower metallicity and to be more
$\alpha$-enhanced than the {\it in situ} disc population.  {\it In
  situ} disc populations also tend to have lower velocity dispersions
than accreted disc populations (see columns 4 and 5 of
Table~\ref{tabdisc}).  This suggests that accreted stars should
contribute primarily to the metal-poor end of a thick disc component
\citep{abadi2003, wyse2010}.

Our central spheroids are populated primarily by old stars which
formed {\it in situ} over a relatively short time interval ($\sim 1.5$
to 4 Gyr; see Table \ref{tabbulge}).  The exceptions are Aq-E-5 and
Aq-F-5, which have been significantly rejuvenated by younger stars, as
indicated by their larger age dispersions.  Generally, accreted stars
in the central spheroids tend to be older, to have formed over very
short intervals, to have lower metallicities and to be more
$\alpha$-enhanced than those formed {\it in situ}.

Our inner stellar haloes are formed by a mixture of {\it in situ} and
accreted stars (Table \ref{tabhalo}). In general, the latter have
similar metallicities but higher $\alpha$-enhancements.  They also
tend to be older and to have formed over a shorter period of
time. Similar trends hold for stars in the outer haloes (Table
\ref{tabhaloex}), although in this case accreted stars tend to be more
metal-rich as well as to have higher $\alpha$-enhancements than the
{\it in situ} population and there is no clear age separation between
the two populations. Apparently, at least a small fraction of outer
halo stars formed {\it in situ} from SNIa enriched material. These may
also have formed at lower redshift, giving rise to the large age
dispersions listed for several galaxies in Table \ref{tabhaloex}.
Recall that stars formed {\it in situ} account for less than 20 per
cent of the stellar mass in the outer haloes and that all bound
satellites have been removed and are considered as separate objects.
Hence, we are here focussing on the ``smooth'' component of stellar
haloes.  Our stellar haloes have a dual formation process since the
ratio of accreted to {\it in situ} stars typically increases from 1
to 5 between the inner and outer haloes (see also \citet{zolo2009}).
We will explore this issue in more detail in the next section.

Old metal-rich stars can be viewed as chemical fossils
\citep{freeman}.
We have measured the fraction of stars both older than 11 Gyr and with
[Fe/H]$>0$ in each component of each of our simulated galaxies. As
expected the largest fractions are found in the central spheroids,
ranging from eight down to one per cent. The other components have
very few stars with these characteristics (fractions always below
1\%).  This is consistent with the fact that only in deep potential
wells is it possible to reach high levels of enrichment in such a
short period of time.  Although we find no correlation with $F_{\rm
  prog}$, we did detect a trend with the strength of the first
starbursts: those central spheroids with a large fraction of old
metal-rich stars have more than 40 per cent of their total stellar
mass already formed at $z > 3$ (i.e. older than 11 Gyr). Conversely,
the other central spheroids were able to form less than 25 per cent of
their final stellar mass by this time. The lowest fraction of old
metal-rich stars is found in Aq-B-5 which made only 15 per cent of its
final stellar bulge by $ z \sim 3.5$.  Hence, the fraction of old and
metal-rich stars in the central spheroids is linked to how fast star
formation proceeded in the first massive structures and so is of
interest for constraining galaxy formation models, as claimed in
previous work \citep[e.g.][and references therein]{wyse2010}.

\section{The influence of the  masses of disrupted satellites on 
the abundance patterns of accreted components}

\begin{figure}
\hspace*{-0.2cm}\resizebox{4.2cm}{!}{\includegraphics{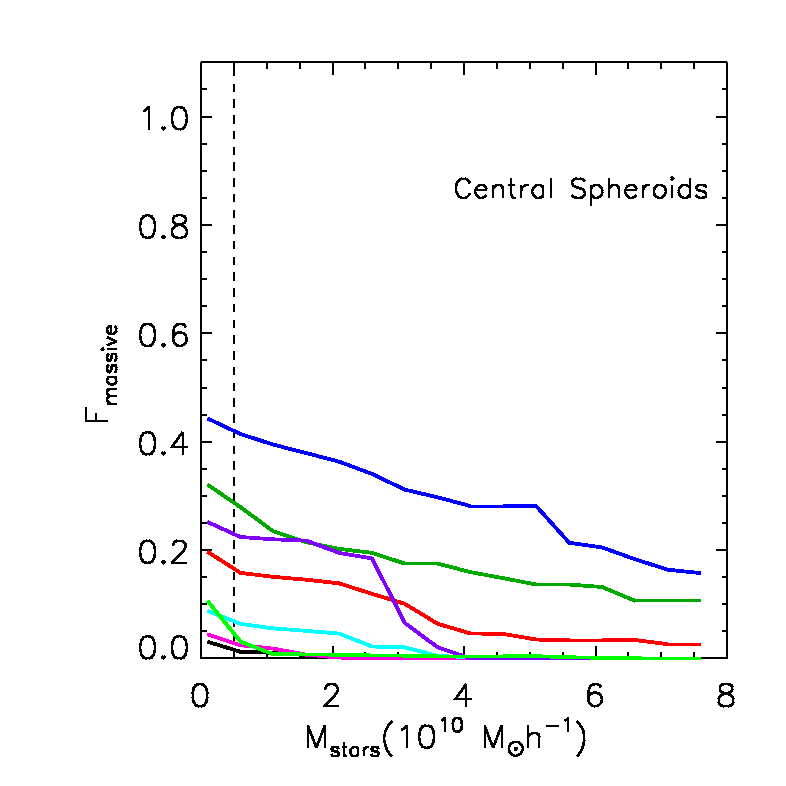}}
\hspace*{-0.3cm}\resizebox{4.2cm}{!}{\includegraphics{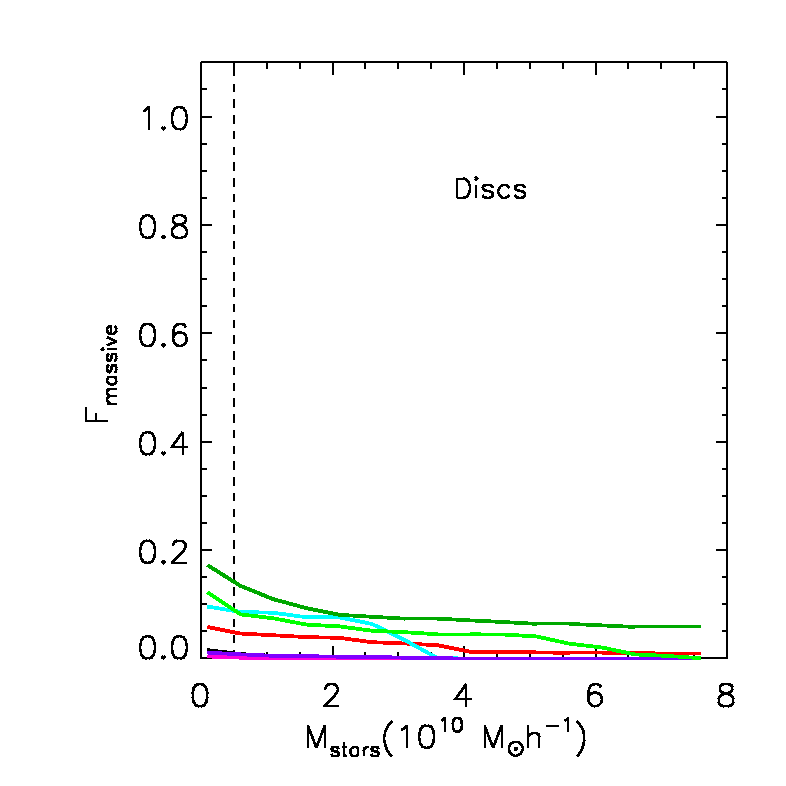}}\\
\hspace*{-0.2cm}\resizebox{4.2cm}{!}{\includegraphics{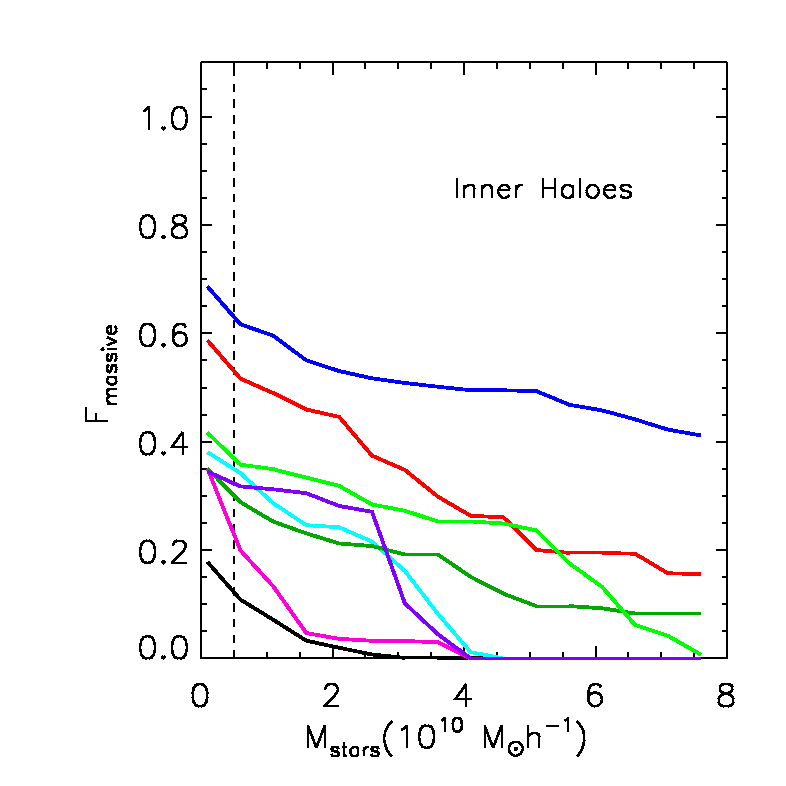}}
\hspace*{-0.3cm}\resizebox{4.2cm}{!}{\includegraphics{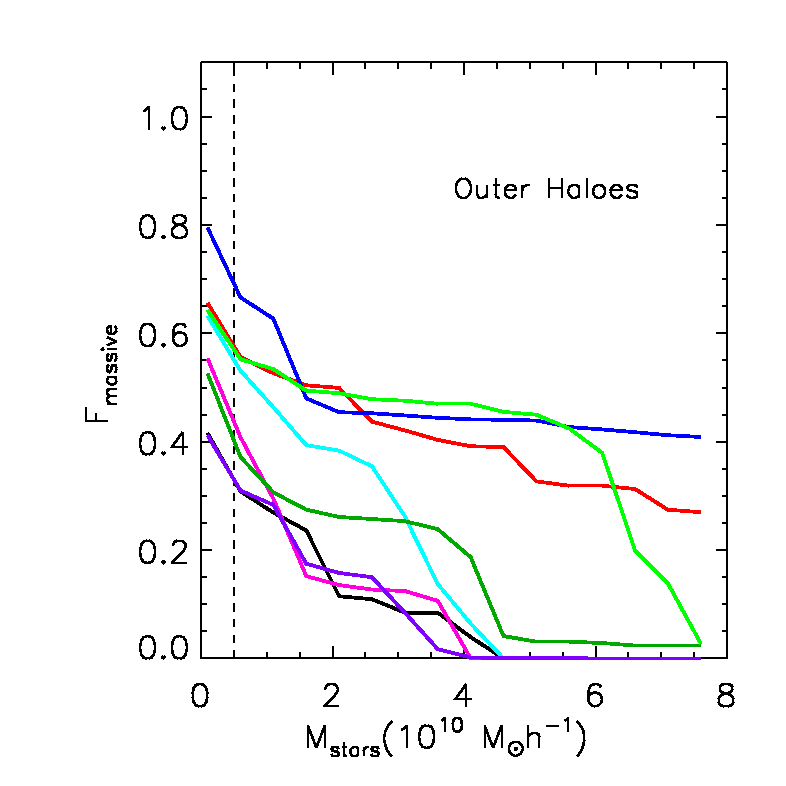}}
\caption{Fraction by mass of each stellar component acquired from satellites
  more massive than a given stellar mass as a function of this stellar
  mass. The dashed lines denote the reference stellar mass, $5 \times 10^9
  M_{\odot}h^{-1}$ See Fig.\ref{fracprog} for the colour code. }
\label{massive}
\end{figure}

The abundance patterns of the various components of our simulated galaxies
depend on the fraction of their stars which was accreted.  However, the
dispersions in these relations are large, suggesting that other factors are
also important.  In particular, the masses of the disrupted satellites vary
from halo to halo \citep[e.g.][]{cooper2010} and these differences could be
reflected in the chemical patterns.  In order to investigate this issue, we
estimate the percentage ($F_{\rm massive}$) of the total stellar mass in a
given dynamical component formed in satellites with stellar masses larger than
a certain value. As can be seen in Fig.~\ref{massive}, the stellar mass of
contributing satellites varies widely. The variation is particularly marked
for the inner and outer stellar haloes which were assembled primarily from
satellite debris.  As a reference value for $F_{\rm massive}$ we take that
measured for stellar mass $ 5 \times 10^9 M_{\odot}h^{-1}$, similar to the
stellar masses of the surviving satellites at $z=0$ in the Aquarius
simulations\footnote{Note that we do not use the ratio between the mass of the
  progenitor and that of the satellite because metallicities are related to
  the total stellar mass which could have enriched the system and to its
  potential well which affects the overall star formation efficiency and the
  impact of outflows.}.

\begin{figure}
\resizebox{7cm}{!}{\includegraphics{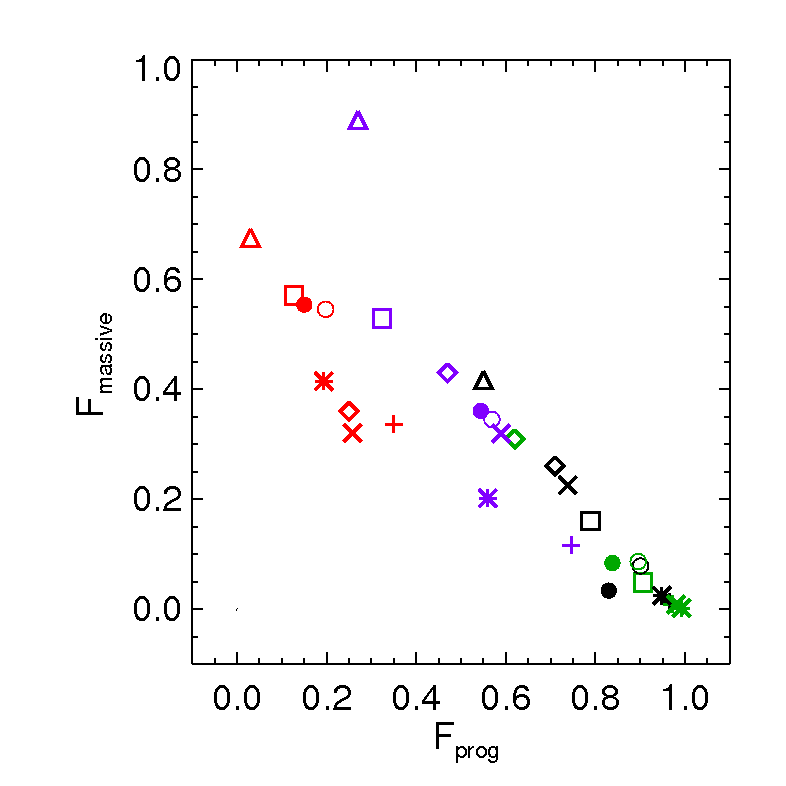}}
\hspace*{-0.2cm}
\caption{Fraction by mass of each stellar components accreted from satellites
  with stellar masses larger than $5 \times 10^9 M_{\odot}h^{-1}$ ($F_{\rm
    massive}$) as a function of $F_{\rm prog}$. See Fig.\ref{medianas} for the
  symbol code. }
\label{fracsat}
\end{figure}

In Fig.~\ref{fracsat}, we show the mass fraction of stars formed in satellites
more massive than $ 5 \times 10^9 M_{\odot}h^{-1}$ ($F_{\rm massive}$) as a
function of $F_{\rm prog}$. There is a clear anti-correlation which simply
reflects the fact that the larger the mass fraction formed in satellites, the
larger the chance that massive objects have contributed.  In the discs and
central spheroids, which formed primarily {\it in situ}, the accreted stars
come mainly from small satellites (i.e. at most $\sim 20$ per cent from
massive ones).  The central spheroids of Aq-E-5 and Aq-F-5 show the largest
accreted fractions and a substantial fraction of their stars come from massive
satellites, $ \sim 0.30 - 0.40$ per cent, consistent with the fact that they
alsos show the lowest median [O/Fe] (Fig.\ref{medianas}).  The outer and the
inner haloes have the largest contributions from stars accreted from massive
satellites while the lowest such contributions are found in discs. As we show
in Fig.~\ref{massive}, the mass functions of the contributing satellites vary
significantly from galaxy to galaxy, leaving their imprints on the chemical
abundances.

\begin{figure}
\hspace*{1cm}\resizebox{7cm}{!}{\includegraphics{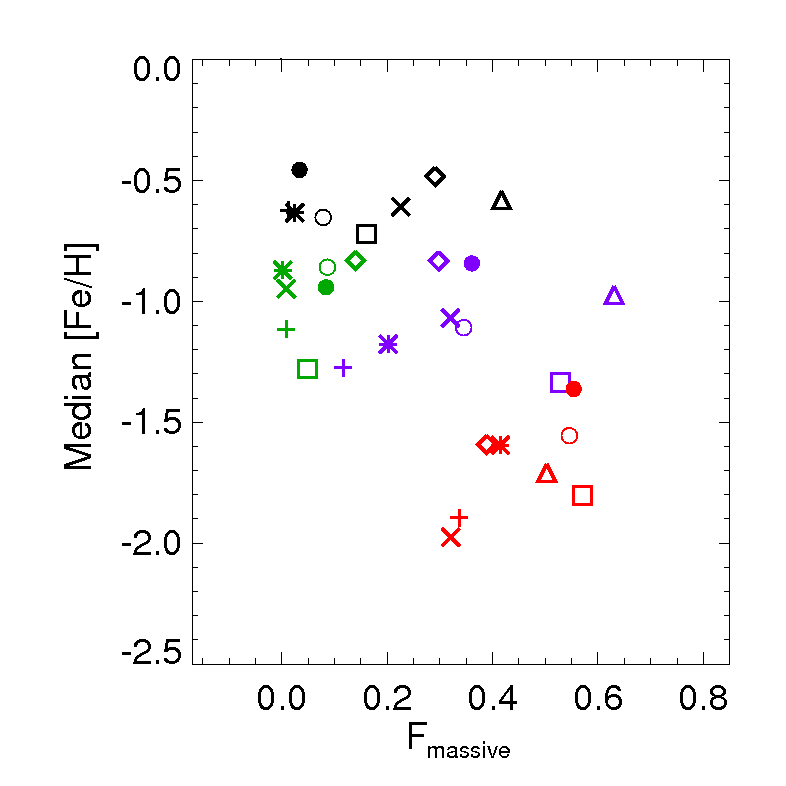}}
\hspace*{1cm}\resizebox{7cm}{!}{\includegraphics{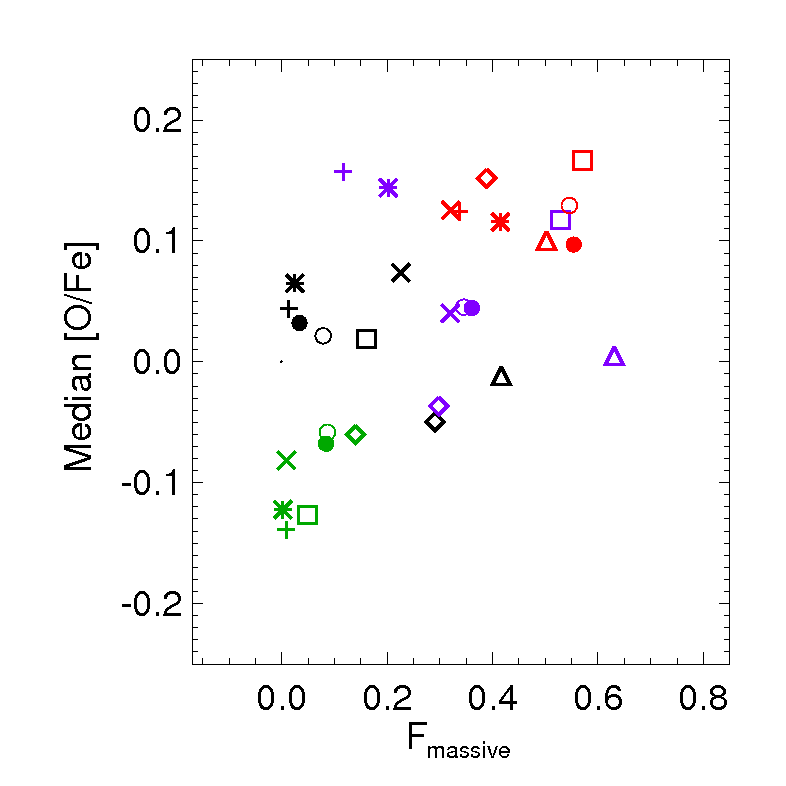}}
\hspace*{-0.2cm}
\caption{Median [Fe/H] and median [O/Fe] as a function of $F_{\rm massive}$,
  the mass fraction of stars accreted from satellites with virial masses
  larger than $5 \times 10^9 M_{\odot}h^{-1}$.  See Fig.\ref{medianas} for
  symbol code.}
\label{mediamassive}
\end{figure}

In Fig. ~\ref{mediamassive}, we show the [Fe/H] and [O/Fe] medians as a
function of $F_{\rm massive}$.  The stellar populations of the inner and outer
haloes tend to be more metal-rich when larger fractions of their stars are
accreted from massive satellites (see Fig. ~\ref{mediamassive}, upper panel).
In this case, the correlation is clearer than that with the fraction of {\it
  in situ} stars.  The inner haloes of Aq-B-5 and Aq-F-5 show similar levels
of enrichment to their outer haloes.  From Fig.~\ref{massive} we see that the
outer and inner haloes of these particular galaxies have similarly important
contributions from stars formed in massive systems, while for the other
galaxies, the contribution of stars acquired from massive satellites decreases
significantly as one moves from the inner to the outer stellar halo.  Thus,
the stellar masses of the satellites where accreted stars were formed is an
important factor in determining their mean level of enrichment: more
metal-rich halo stars tend to come from more massive satellites.

As shown in Fig. ~\ref{mediamassive} (lower panel), $\alpha$-enhancement tends
to increase with a decreasing contribution from massive satellites for all
three spheroidal components. The lowest median [O/Fe] values are found in
central spheroids. Inner haloes extend over a larger range of both [O/Fe] and
$F_{\rm massive}$, while outer stellar haloes show the highest
$\alpha$-enhancements, reflecting the fact that they are dominated by old
stars formed during short starbursts.

A dual origin for halo stars is suggested by observations of the Milky Way
which show the inner halo to have a stellar metallicity distribution peaking
at [Fe/H]$\approx -1.6$ while the outer halo is dominated by less enriched
stars with peak [Fe/H]$\approx -2.2$ \citep{caro2007}.  There are also new
observations suggesting differences in kinematics and $\alpha$-enrichment for
stars in the Milky Way stellar halo  
which might indicate the presence of two distinct stellar populations \citep[e.g.][]{niss2010, beers2011}.
   We now explore
if the relative abundances of the inner and outer stellar field haloes of the Aquarius
systems are comparable to these observed values and if a radial trend is a
common characteristic of all our simulated systems.  We stress the fact that
in this section, we are analysing the diffuse stellar population in each component.
 Surviving satellites are explored in the last  section of this paper.

\begin{figure}
\hspace*{.1cm}\resizebox{7cm}{!}{\includegraphics{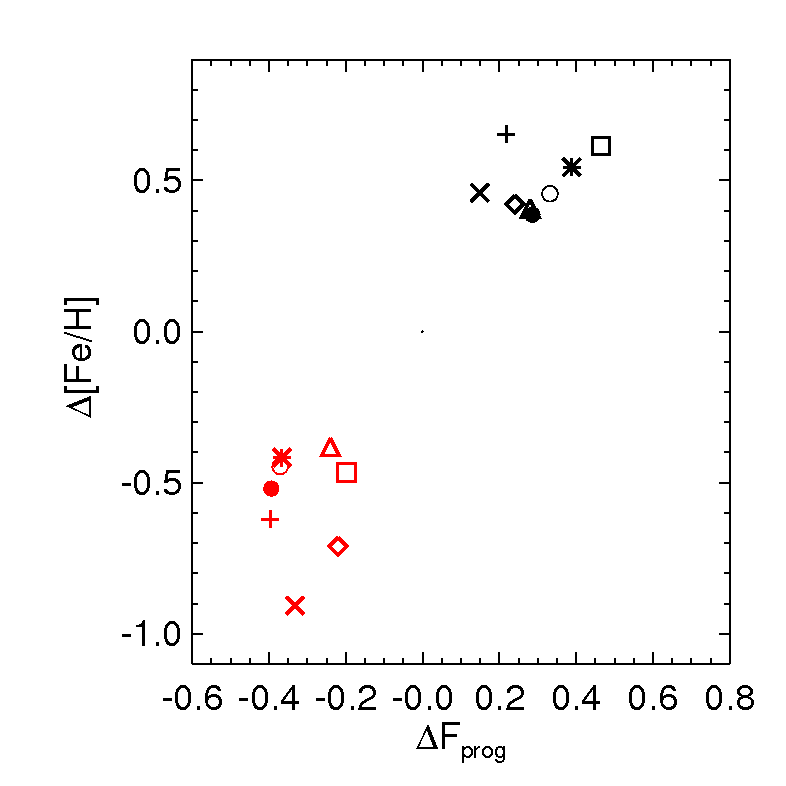}}
\hspace*{.1cm}\resizebox{7cm}{!}{\includegraphics{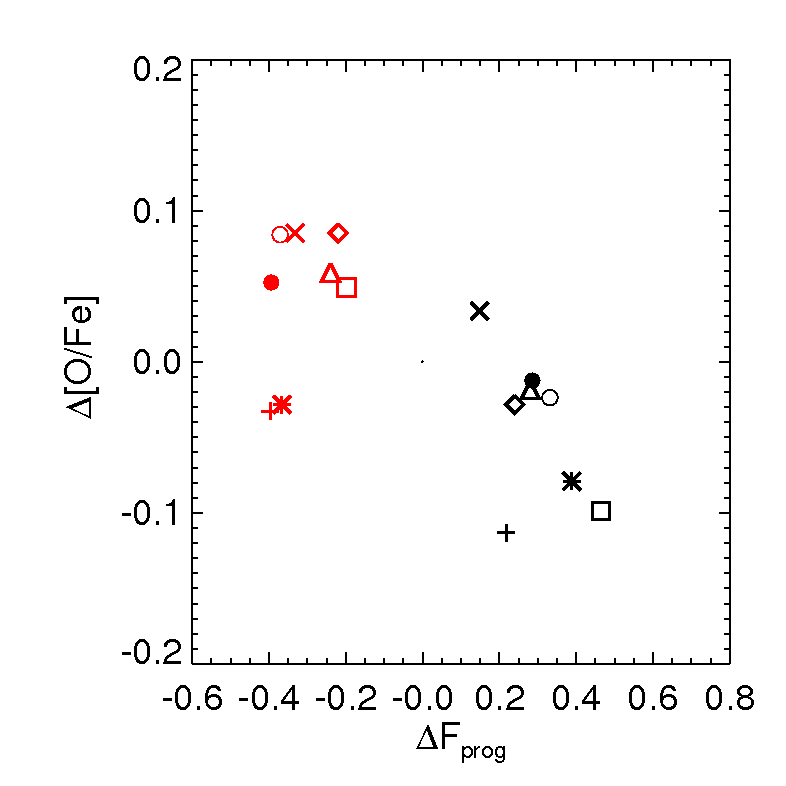}}
\hspace*{-0.2cm}
\caption{ [Fe/H] and [O/Fe] abundances of central spheroids relative to inner
  stellar haloes (black symbols) and of outer stellar haloes relative to inner
  haloes (red symbols) haloes as a function of the differences between the
  fractions of stellar mass formed {\it in situ} ($\Delta{\rm F_{\rm prog}}
  $). See Fig. ~\ref{medianas} for the symbol code.}
\label{diff}
\end{figure}

We have estimated the differences in median [Fe/H] between the inner and outer
stellar haloes of each of our eight simulated galaxies and looked for
relations to the fraction of {\it in situ} stars.  In all systems, the outer
halo is less chemically enriched than the inner one, with differences varying
from $\sim -0.6$ dex to $\sim -1.0$ dex (Fig. ~\ref{diff}).  The values quoted
by \citet{caro2007} for the Milky Way imply a difference of $ \sim -0.6$
dex, consistent with the range in the simulations where these gradients
reflect the gradients in assembly history.  \citet{zolo2009} and
\citet{zolo2010} reached similar conclusions, but used a different approach,
so caution is needed in any detailed comparison with our results.  The
dispersion in chemical gradients reflects a variety of factors, including the
accreted mass fraction, the mass function of disrupted satellites, the ages of
their stars, the mixing effiency, and the strength of outflows. For example,
Aq-G-5 has the largest chemical differences between the inner and outer haloes
due to a larger contribution of debris from massive satellites to the inner
halo than to the outer one.  $\alpha$-enhancements depend principally on the
star formation timescales and the level of chemical mixing both in disrupted
satellites and in the main progenitor systems. The inner haloes of our
Aquarius galaxies are typically $\sim 0.8$ dex less $\alpha$-enhanced than
their outer haloes, but in Aq-A-5 and Aq-C-5 the inner haloes are slightly
more $\alpha$-enriched than their outer components for given $\Delta FÌ£_{\rm
  prog}$.  Fig. ~\ref{massive} shows that these two inner haloes obtained
their accreted stars from much smaller satellites than the corresponding outer
haloes, contrasting with the other Aquarius systems.

A similar comparison can be done for the central spheroids and the inner
haloes.  In this case we found metallicity differences ranging from 0.9 to
0.4~dex with central spheroids always more enriched than the inner haloes.
Based on observations reported in the literature \citep[e.g][]{wyse2010}, we
estimate a corresponding difference of  $\sim 0.35$ in the Milky Way, in reasonable
agreement with our results even though none of our simulated systems has a
morphology resembling that of the Milky-Way.  These trends are shown in
Fig.\ref{diff}.  Larger differences in metallicity, are again associated with
larger differences in formation history between the two components.  The
central spheroids are also systematically less $\alpha$-enriched than the
inner haloes, and again this trend is more important for larger differences in
the way the components were assembled.

\section{Surviving satellites}

The chemical abundances of Milky Way satellites and nearby dwarf galaxies have
been a subject of much research. Observations suggest significant differences
between the abundance patterns of stars in the Galactic halo and those in
dwarf galaxies: stars in dwarfs tend to have lower [$\alpha /$Fe] than similar
metallicity stars in the Galactic halo \citep[e.g.][]{tols03} even at low
[Fe/H] levels.  Thus it appears that the Galactic halo could not have been
built from systems similar to the current satellites.  In this Section, we
compare these two populations in our simulated systems.  For this analysis, we
have only considered satellites represented by more than 2000 total particles. 
 Two of our galaxies
(Aq-E-5 and Aq-H-5) are excluded from this analysis, since they have no
satellites satisfying this numerical requirement.

\begin{figure*}
\hspace{-0.2cm}\resizebox{5.5cm}{!}{\includegraphics{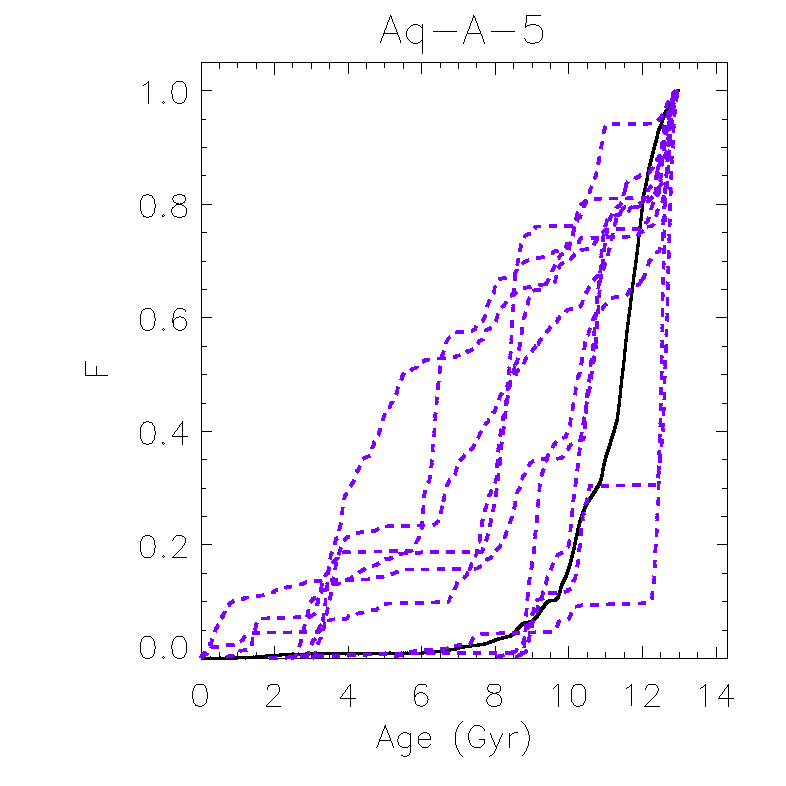}}
\hspace{-0.5cm}\resizebox{5.5cm}{!}{\includegraphics{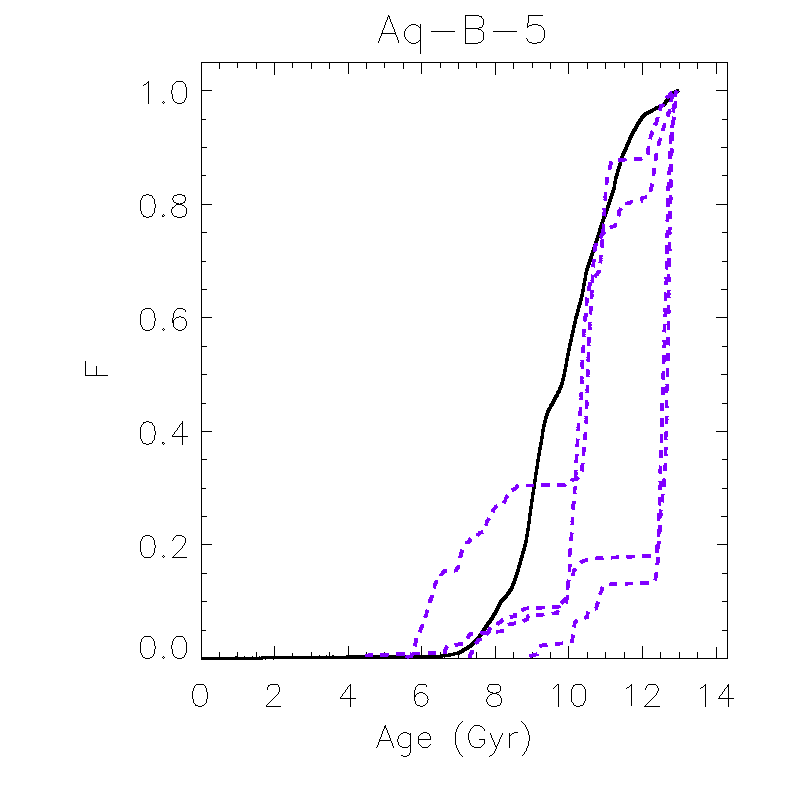}}
\hspace{-0.5cm}\resizebox{5.5cm}{!}{\includegraphics{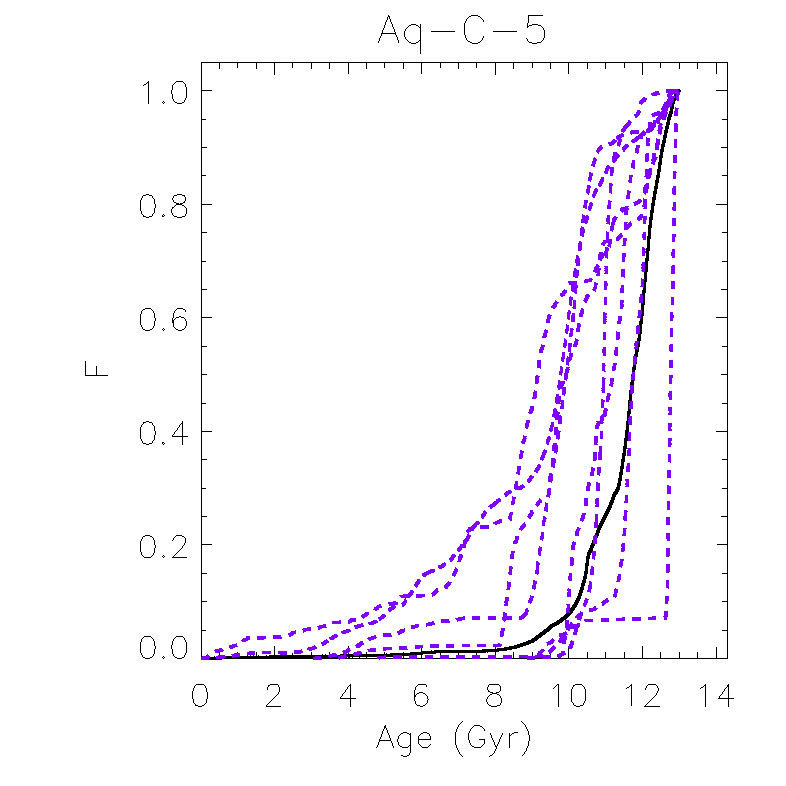}}\\
\hspace{-0.2cm}\resizebox{5.5cm}{!}{\includegraphics{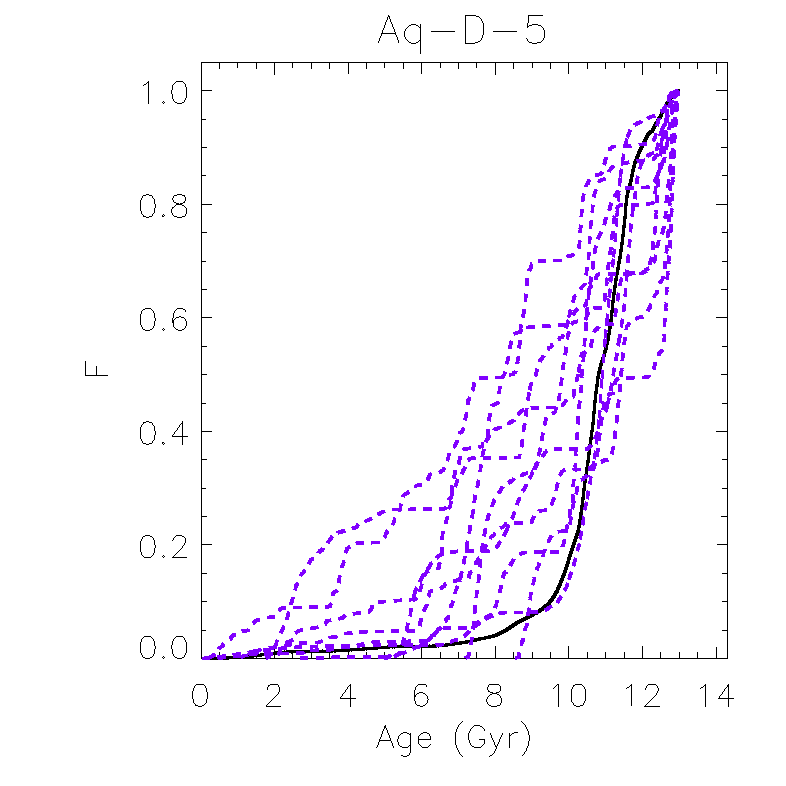}}
\hspace{-0.5cm}\resizebox{5.5cm}{!}{\includegraphics{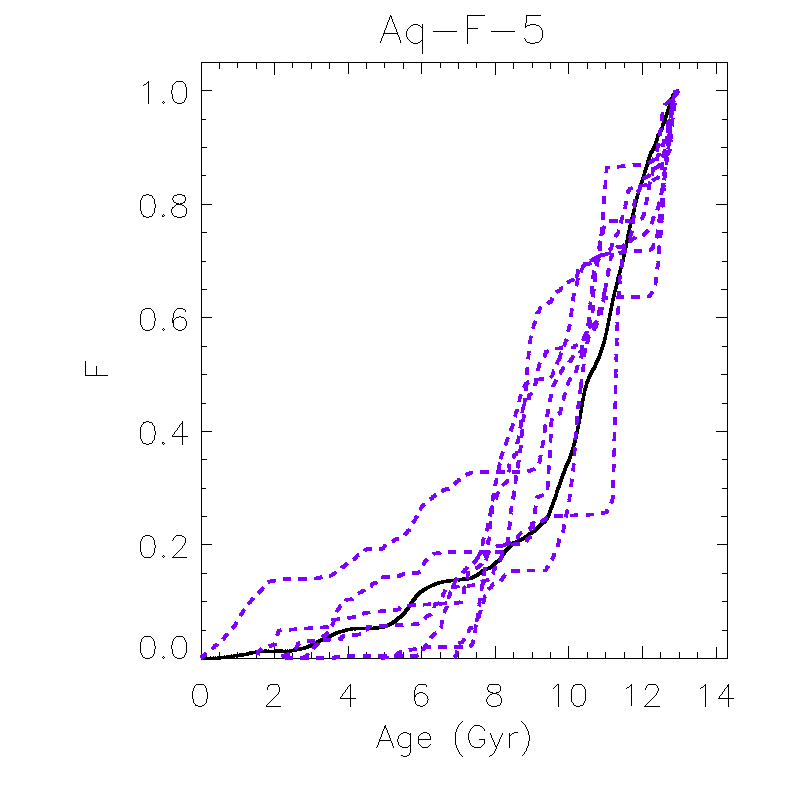}}
\hspace{-0.5cm}\resizebox{5.5cm}{!}{\includegraphics{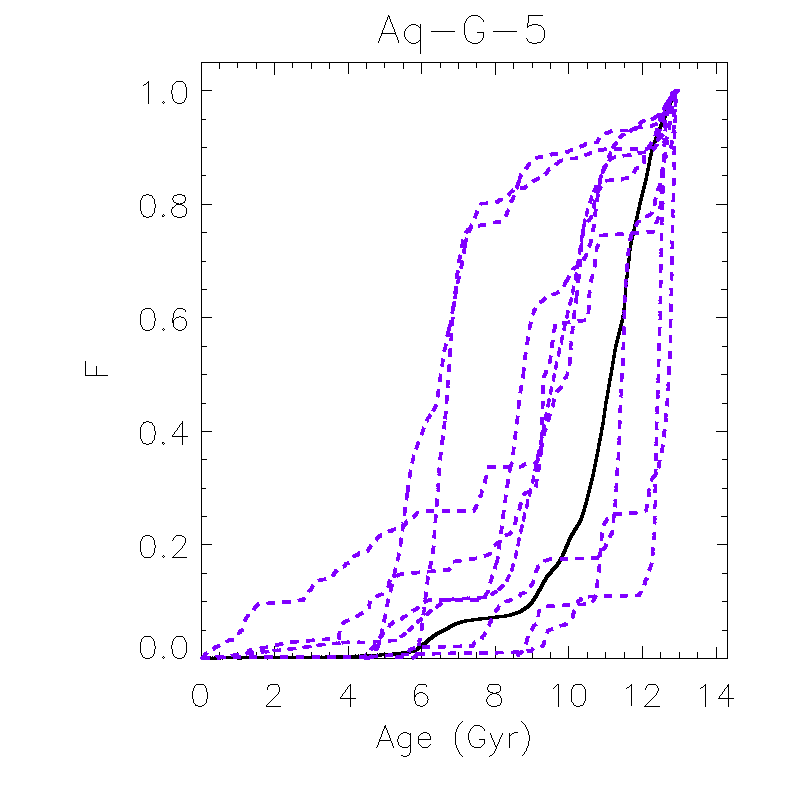}}
\hspace*{-0.2cm}
\caption{Cumulative stellar mass as a function of age for surviving satellites
  (violet dashed lines) and for the outer stellar haloes (black solid lines)
  of six of our Aquarius galaxies.  Aq-E-5 and Aq-H-5 have only very low-mass
  satellites and have not been included in this plot.  }
\label{sfrsatelites}
\end{figure*}

In Fig. ~\ref{sfrsatelites}, we show cumulative stellar mass as a function of
stellar age for each satellite of the remaining six galaxies. Clearly, star
formation histories differ considerably, with most showing several starburst
episodes.  For comparison, we have included the star formation histories of
the outer haloes of the corresponding systems.  Most surviving satellites have
experienced several starbursts and are dominated by younger stellar
populations than the corresponding outer haloes.  Few satellites have
significant old stellar populations and only one halo, Aq-B-5, has satellites
dominated by older stars than those in the corresponding outer halo.  This
figure shows the variety in the star formation histories of satellites
inhabiting similar haloes, illustrating the difficulty of comparing the Milky
Way to any specific simulation.  For this reason, we compare the chemical
properties of accreted stars in our four stellar components to those of stars
in the surviving satellites on a statistical basis.

\begin{figure}
\resizebox{7cm}{!}{\includegraphics{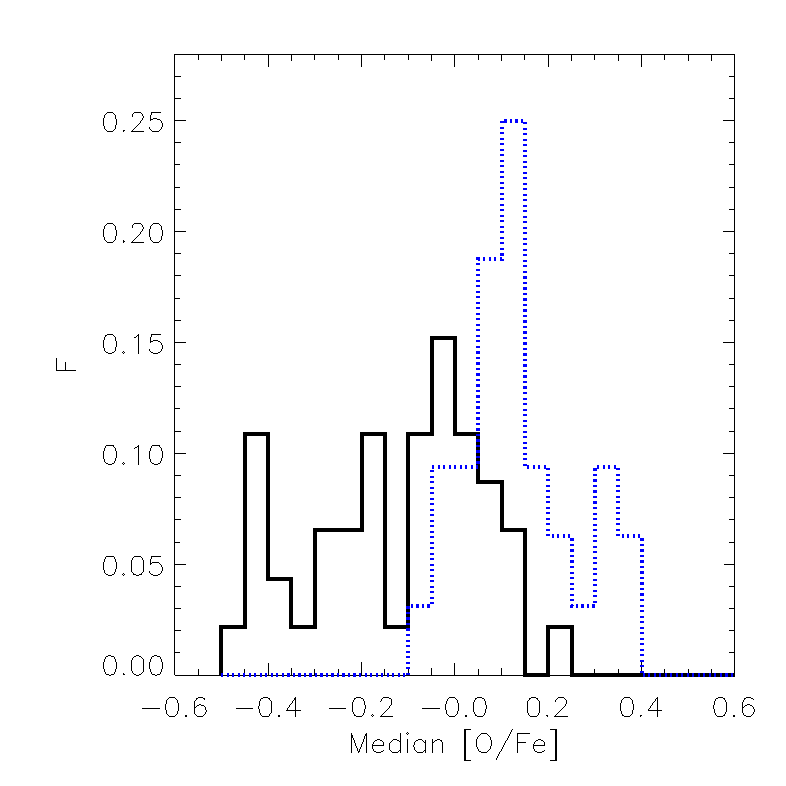}}
\resizebox{7cm}{!}{\includegraphics{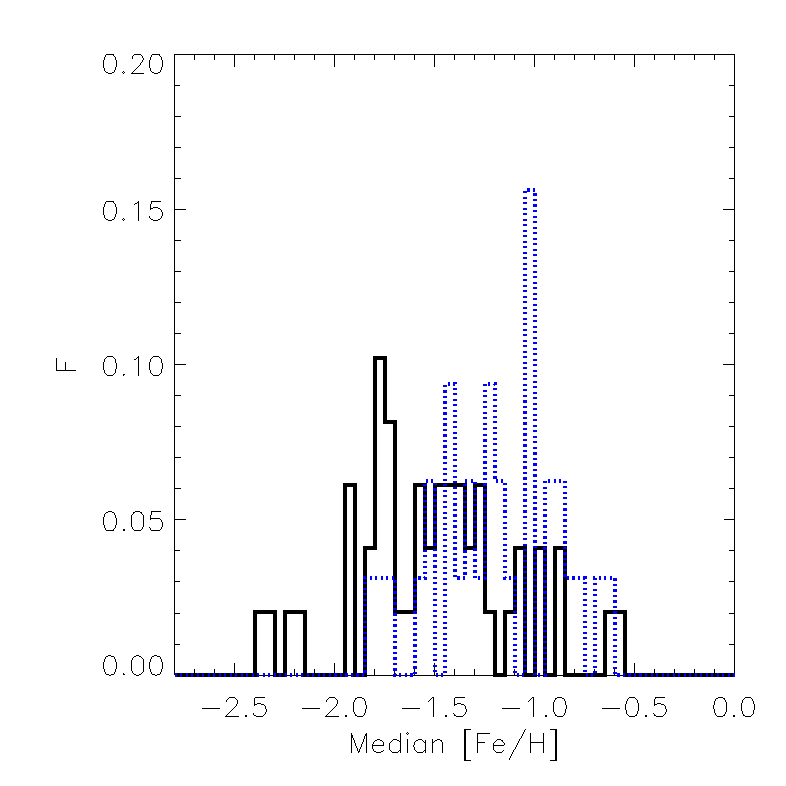}}
\hspace*{-0.2cm}
\caption{Distributions of median [O/Fe] (upper panel) and of median [Fe/H]
  (lower panel) for the accreted component of outer haloes (dotted lines)
  and for stars in surviving satellites (solid lines). }
\label{satsathist}
\end{figure}

Fig. ~\ref{satsathist} shows the distributions of median [O/Fe] and median
[Fe/H] for stars in surviving satellites (solid lines) and for accreted stars
in the outer haloes (dotted lines).   Accreted stars tend to be
$\alpha$-enhanced, reflecting the fact that they formed quickly and in short
bursts, so that SNII enriched them but SNIa exploded too late to affect their
abundances.  In contrast, surviving satellites contain significantly more
weakly $\alpha$-enhanced stars at all metallicities. This is a consequence of
their extended star formation histories (Fig. ~\ref{sfrsatelites}). As shown
in previous papers, our feedback model successfully drives mass-loaded
galactic winds out of star-forming galaxies with a wide range of masses, and
with a specific mass-loss rate which depends on the potential well depth of
the systems. These winds regulate star formation activity and determine the
fraction of available baryons which are turned into stars
\citep{scan08,scan09,scan10,sawala2010,dero10,sawala2011}. As a result, the
chemical properties and ages of the stars in surviving satellites are
statistically different in our simulations from those of stars acquired by
satellite disruption.

An interesting point of comparison between our current model and observation
concerns the percentage of very metal-poor stars ([Fe/H] $<-3$) in the
surviving satellites. In our simulations this varies from 0.5 to 20 per cent
(with an average of $\sim 10$ per cent), depending on details of individual
star formation histories. These stars are $\alpha$-enhanced and old.  The
apparent lack of such low-metallicity stars in real systems has been suggested
as a possible problem for hierarchical galaxy formation models in the past \citep[e.g.][]{shetr2001,helm06}.  However, more recent studies have discovered extremely
metal-poor stars in dwarf galaxies \citep{kirb08,kirb09,norri08, norri2010}.
These new observations, together with improved metallicity indicators
\citep{stark2010}, are causing a revision of ideas about the low metallicity
tail of the stellar populations, which are clearly more extensive than
previously thought. These tails are of great interest for investigations of
the possible impact of other processes such as pre-enrichment by Population
III stars which might have led to a floor for Population II metallicities,
leading to a deficit of extremely metal-poor stars in dwarfs.  Improved
observational surveys of low metallicity stars in nearby satellites are thus
very much relevant for constraining galaxy formation models.

Our results are encouraging in that they show differences between the chemical
properties of accreted halo stars and those of stars in surviving satellites
which are quite similar to the trends observed in the Milky Way.  Higher
resolution simulations are clearly needed to confirm these findings, but they
already suggest that these trends are not only compatible with the standard
model of hierarchical formation in a $\Lambda$CDM cosmology, but may, indeed,
provide direct evidence for the assembly paths it predicts.

\section{Conclusions}

In this paper we study the chemical patterns of the stellar populations in
eight Milky-Way mass-sized galaxies simulated within the $\Lambda$CDM
scenario. Our version of {\small GADGET-3} includes the modelization of SN
feedback describing the release of energy and chemical elements in the
interstellar medium so that the chemical enrichment of baryons can be followed
consistently as galaxies are assembled.  This is the largest sample of high
resolution Milky-Way mass-sized galaxies so far run with a physical motivated
SN model fact that provides us with the unique possibility to distinguish
those chemical properties which are common to all galaxies from those that are
determined by the particular history of assembly of each galaxy.

Our results can be summarized as follows

\begin{itemize}
\item Discs: The surviving discs show very similar chemical properties.  They
  are more chemically enriched than the stellar halo but slightly less
  enriched than the central spheroids.  The stars in discs show the lowest
  levels of $\alpha$-enhancement. This is  consistent with being formed from SNIa
  pre-enriched material. These stars formed mainly {\it in situ}.  Most of the stars in the discs formed from material which has never  been 
   part of a galactic wind. 
 In general, the discs formed inside-out as shown by
  \citet{scan09,scan11}, and during this process different starbursts enriched
  continuously the gas.  We identified that up to 15 per cent of the
  stars in the discs formed in satellite systems confirming previous results
  \citep{abadi2003}. These stars tend to older and formed in shorter
  starbursts in comparison to the dominating stellar population formed {\it in
    situ}.  Accreted stars have systematically higher $\alpha$-enhancement and
  are more supported by velocity dispersion. These stars could contribute to a
  thick disc component.  Most of the accreted stars come from satellites
  smaller than $5\times 10^9 {\rm M _{\odot}}h^{-1}$ total stellar mass.

\item Central spheroids: They are formed mainly by old stars, born in
  starbursts with different duration ranging from 2 to 4 Gyrs
  approximately. The stellar populations in central spheroids have been formed
  {\it in situ} with contributions of less than $\sim 20$ per cent of stars
  formed in satellites. Two systems out of eight (Aq-E-5 and Aq-F-5), show
  larger contributions of accreted stars. These central spheroids have been
  also recently rejuvenated by news stars formed mainly {\it in situ} by the
  infall of enriched gas(mean stellar age of the new population 4-5 Gyrs).
  This can be confirmed by the lower $\alpha$-enhancement showed by these two
  central spheroids with respect to the rest of the systems.
Very old and metal-rich stars are principally located in the central spheroids
and their frequency can be correlated with rate of star formation in the
earliest epoch of formation.
 
\item The inner and outer field haloes: As one moves outward the contribution
  from disrupted stars to the field stellar haloes increases dramatically. The
  inner haloes have contribution from both kind of stars, but the outer haloes
  have been mainly formed by accreted stars.  The mean metallicity shows a
  large dispersion which correlates with the fraction of stars acquired from
  massive satellites (i.e. larger than $5\times 10^9 {\rm M _{\odot}}h^{-1}$)
  so that the larger this fraction, the larger the mean level of enrichment
  of the halo.  This is valid for both the inner and the outer haloes.

We find that in all simulated stellar haloes the stars in the outer region
are systematically less enriched.  The differences in the chemical properties
between the inner and the outer field haloes are not only the consequence of
the different relative contributions of {\it in situ} and accreted stars
\citep{zolo2009} but also reflect the mass function of the satellites which
contributed with stars to form them.

The level of $\alpha$-enhancement is higher in outer field stellar haloes
reflecting the fact that they have been mainly assembled from old stars formed
in smaller structures, in single short-duration starbursts.  Only stars
accreted from very massive systems tend to be, on average, less
$\alpha$-enriched at a given metallicity than the global population of the
halo \citep{zolo2010}.  
The masses of the specific satellites where the accreted
stars were formed are reflected in the chemical properties of the
inner and outer haloes, imprinting features which distinguish one
system from another.

\item Surviving satellites: Each simulated Milky-Way mass galaxy has a system
  of surviving satellites with particular characteristics.  The surviving
  satellites show a variety of star formation histories but in general, they
  formed stars in a series of starburst episodes which extended even to the
  present time.  As a consequence, their chemical patterns differ from those
  of the stars that contributed to the formation of the stellar components of
  the main galaxies. On average, stars in surviving satellites are younger,
  low metallicity and low $\alpha$-enriched.
\end{itemize}

Although there are many open questions and issues to be solved regarding
galaxy formation, particularly those that could be related to the formation
and surviving of disc galaxies, it is encouraging that the global chemical
trends observed in nearby galaxies can be obtained in $\Lambda$CDM
scenarios. Our work shows the importance of studying chemical evolution to
understand galaxy formation and the relevance of doing so within a
cosmological context.

\section*{Acknowledgements}
We thank the referee, Brad Gibson, for his useful comments which helped to
improve this paper.
PBT thanks Manuela Zocalli, Patricia Sanchez-Blazquez and Tim Beers for  their suggestions and comments 
 and the ISIMA at University of Santa Cruz for the stimulating environment that
helped to put this paper together. 
This work was partially funded by PROALAR 07 (DAAD-Secyt collaboration), PICT 32343 (2005) and
PICT Max Planck 245 (2006) of the Ministry of Science and Technology (Argentina).

\end{document}